\documentclass[12pt,a4paper]{article}
\usepackage{a4wide}
\usepackage{amsmath}
\usepackage{amssymb} 
\usepackage{epsfig}
\usepackage{subfigure}
\usepackage{exscale}
\usepackage{float}
\usepackage{bbm}
\usepackage[numbers,sort&compress]{natbib}
\usepackage{pst-plot, pstricks,pst-math}
\usepackage{fancybox,amssymb,color}
\usepackage{graphicx}
\usepackage{pstricks, color, graphicx, epsfig, psfrag}
\usepackage{amsfonts,amsmath,amssymb,slashed}
\usepackage{dsfont}
\usepackage{bbm,bm}
\usepackage{fancyhdr, a4wide}
\usepackage[english]{babel}
\usepackage{subfigure}

\makeatletter
\@addtoreset{equation}{section}
\makeatother

\title{Uniform Perpendicular and Parallel Staggered Susceptibility of Antiferromagnetic Films}

\author{Christoph P.\ Hofmann \\ \\
Facultad de Ciencias, Universidad de Colima \\
Colima C.P.\ 28045, Mexico}

\begin{document}

\maketitle

\begin{abstract} \normalsize

We investigate the thermodynamic behavior of antiferromagnetic films in external magnetic fields oriented perpendicular to the staggered
magnetization. Within the systematic effective Lagrangian framework we first calculate the two-point function and the dispersion relation
for the two types of magnons up to one-loop order. This allows us to split the two-loop free energy density into a piece that originates
from noninteracting dressed magnons, and a second piece that corresponds to the genuine magnon-magnon interaction. We then discuss the
low-temperature series for various thermodynamic quantities, including the parallel staggered and uniform perpendicular susceptibilities,
and analyze the role of the spin-wave interaction at finite temperature.

\end{abstract}
\maketitle

\section{Motivation}
\label{Intro}

The behavior of antiferromagnetic films in an external magnetic field has been studied by many authors \citep{AUW77,FP85,Fis89,Sin89,WOH91,
FKLM92,HWA92,Glu93,MG94,SSS94,WLW94,ZN98a,ZN98b,San99,SM01,SVB01,Yun02,SS02,HSR04,VCC05,Che05,SSKPWLB05,JBMD06,HSSK07,YK07,KHK07,KSHK08,
CZ09,LL09,FZSR09,HRO10,SST11,SST13,PR15}. Still, a fully systematic effective Lagrangian field theory investigation of antiferromagnetic
films subjected to mutually orthogonal magnetic and staggered fields, has only been performed very recently \citep{Hof17}. In the present
study we further pursue this endeavor by evaluating the two-point function and the dispersion relation for the two types of magnons. Using
these dressed magnons as fundamental degrees of freedom, we then calculate the parallel staggered and uniform perpendicular
susceptibilities at finite as well as zero temperature, and compare our results with the literature.

The effective Lagrangian method, originally developed in particle physics \citep{GL84,GL85}, has been transferred to condensed matter
systems in Refs.~\citep{Leu94a,ABHV14}. At low temperatures the only relevant degrees of freedom are the spin waves (magnons) that can be
interpreted as the Goldstone bosons of the spontaneously broken internal spin symmetry $O(3) \to O(2)$.\footnote{It should be noted that
the Mermin-Wagner theorem excludes spontaneous symmetry breaking in two-dimensional isotropic Heisenberg antiferromagnets at finite
temperature. Still, the magnons dominate the low-temperature properties of antiferromagnetic films, and the staggered magnetization takes
nonzero values at low temperatures and weak external fields. In the present study we hence use the terms "order parameter" and "staggered
magnetization" synonymously.} The effective Lagrangian can be constructed systematically and the thermodynamic properties of the system can
be obtained order by order in the low-energy (low-temperature) expansion.

Within the effective field theory framework, the partition function for antiferromagnetic films in magnetic fields oriented perpendicular
to the staggered magnetization, has been derived in Ref.~\citep{Hof17}. In that article, finite-temperature series for the pressure, order
parameter and the magnetization have been presented and the effect of the spin-wave interaction in these thermodynamic quantities has been
discussed. The present work goes beyond that reference in various ways. First of all, the parallel staggered and uniform perpendicular
susceptibilities have not been addressed in Ref.~\citep{Hof17} within effective field theory, and neither have they been explored before in
the literature in the regime of weak magnetic and staggered fields. Second, here we calculate the two-point function and extract the
dispersion relation for the dressed magnons which allows for a clear-cut definition of the spin-wave interaction part in thermodynamic
quantities. Moreover, we investigate the impact of the magnon-magnon interaction in the susceptibilities and revisit -- using the dressed
magnon picture -- the role of the interaction in the pressure, staggered magnetization and magnetization. Overall, the effects of the
spin-wave interaction at finite temperature are rather subtle and only play a minor role. We also consider a hydrodynamic relation obeyed
by antiferromagnetic films at zero temperature, and extract the numerical values for the uniform perpendicular susceptibility in zero
external fields both for the square and honeycomb lattice Heisenberg antiferromagnet.

The article is structured as follows. The low-temperature representation for the two-loop free energy density is briefly reviewed in
Sec.~\ref{preliminaries} to set the basis for the subsequent analysis. The two-point function for the two types of magnons is calculated in
Sec.~\ref{TwoPointFunction} where we also provide the dispersion relations for the dressed magnons in presence of mutually orthogonal
magnetic and staggered fields. In Sec.~\ref{LowTSeries}, after a few remarks on the relevant energy scales, low-temperature series for the
pressure, order parameter, magnetization -- as well as for the parallel staggered and uniform perpendicular susceptibilities -- are
discussed. A central theme is how the genuine magnon-magnon interaction manifests itself at finite temperature. Finally, in
Sec.~\ref{conclusions} we conclude.

\section{Preliminaries}
\label{preliminaries}

The starting point in the microscopic analysis of antiferromagnets subjected to magnetic ($\vec H$) and staggered ($\vec H_s$) fields, is
the Hamiltonian
\begin{equation}
\label{microAF}
{\cal H} = - \, J \, \sum_{n.n.} {\vec S}_m \! \cdot {\vec S}_n - \sum_n {\vec S}_n \cdot {\vec H} - \sum_n (-1)^n {\vec S}_n \! \cdot
{\vec H_s} \, , \qquad J = const.
\end{equation}
Here $J < 0$ is the exchange integral and the symbol "n.n." indicates that the summation involves nearest neighbor spins only.\footnote{We
assume that the two-dimensional lattice is bipartite. In Sec.~\ref{LowTSeries} we explicitly refer to the square and honeycomb lattice.}
The first term is invariant under O(3) spin rotations, whereas the other terms are not. But if we assume that the magnetic and staggered
fields are weak, the rotation symmetry is still approximate -- this is the situation we consider in the present study. The crucial
observation is that the ground state of the antiferromagnet is only invariant under O(2): the (approximate) spin symmetry O(3) is
spontaneously broken. Accordingly, two gapped spin-wave excitations ("pseudo-Goldstone bosons" in particle physics jargon) emerge in the
low-energy spectrum and the systematic effective Lagrangian field theory can be put to work.

It is not our intention, however, to describe the effective field theory method in much detail here. The interested reader is referred to the
pioneering article by Leutwyler, Ref.~\citep{Leu94a}, as well as Secs. IX-XI of Ref.~\citep{Hof99a}. Also, in the preceding paper,
Ref.~\citep{Hof17}, a more detailed outline was given in Secs. II and III. Regarding the present investigation, below we just provide the
basic formulas and results.

We refer to the situation where magnetic and staggered fields point in mutually orthogonal directions,
\begin{equation}
\label{externalFields}
{\vec H}_{\perp} = (0,H,0) \, , \qquad {\vec H}_s = (H_s,0,0) \, , \qquad H, H_s > 0 \, .
\end{equation}
Note that ${\vec H}_s$ is aligned with the staggered magnetization that represents the order parameter. The magnons obey the "relativistic"
dispersion relations
\begin{eqnarray}
\label{disprelAFH}
\omega_{I} & = & \sqrt{{\vec k}^2 +  M^2_{I}} \, , \qquad M^2_{I} = \frac{M_s H_s}{\rho_s} + H^2 \, , \nonumber \\
\omega_{I\!I} & = & \sqrt{{\vec k}^2 +M^2_{I\!I}} \, , \qquad M^2_{I\!I} = \frac{M_s H_s}{\rho_s} \, .
\end{eqnarray}
Here $\rho_s$ is the spin stiffness and $M_s$ is the staggered magnetization at zero temperature in the absence of magnetic and staggered
fields. Notice that the magnetic field only affects one of the magnons.

The free energy density $z$ of antiferromagnetic films subjected to mutually orthogonal staggered and magnetic fields, has been derived in
Ref.~\citep{Hof17}:
\begin{eqnarray}
\label{freeEDtwoLoop}
z & = & z_0 - \mbox{$ \frac{1}{2}$} \Big\{ g^{I}_0 + g^{I\!I}_0 \Big\} \nonumber \\
& & + \frac{M_s H_s}{16 \pi \rho^2_s} \, \Bigg\{ \sqrt{\frac{M_s H_s}{\rho_s} + H^2} - \sqrt{\frac{M_s H_s}{\rho_s}} \Bigg\} g^{I}_1
+ \frac{H^2}{4 \pi \rho_s} \, \sqrt{\frac{M_s H_s}{\rho_s} + H^2} \, g^{I}_1  \nonumber \\
& & - \frac{M_s H_s}{16 \pi \rho^2_s} \, \Bigg\{ \sqrt{\frac{M_s H_s}{\rho_s} + H^2} - \sqrt{\frac{M_s H_s}{\rho_s}} \Bigg\} g^{I\!I}_1 \\
& & - \frac{M_s H_s}{8 \rho^2_s} \, \Big\{ {(g^{I}_1)}^2 - 2 g^{I}_1 g^{I\!I}_1 + {(g^{I\!I}_1)}^2 \Big\}
- \frac{H^2}{2 \rho_s}{(g^{I}_1)}^2 + \frac{2}{\rho_s} \, s(\sigma,\sigma_H) \, T^4 \, . \nonumber
\end{eqnarray}
The vacuum energy density $z_0$ that comprises all temperature-independent contributions, reads
\begin{eqnarray}
\label{vacuumEDtwoLoop}
z_0 & = & - M_s H_s - \mbox{$ \frac{1}{2}$} \rho_s H^2
- (k_2 + k_3) \frac{M^2_s H^2_s}{\rho^2_s} - k_1 \frac{M_s H_s}{\rho_s} H^2 -(e_1 + e_2) H^4 \nonumber \\
& & - \frac{1}{12 \pi} \Bigg\{ {\Big( \frac{M_s H_s}{\rho_s} + H^2 \Big) }^{3/2} + {\Big( \frac{M_s H_s}{\rho_s} \Big)}^{3/2} \Bigg\}
- \frac{M_s^2 H_s^2}{64 \pi^2 \rho^3_s} \nonumber \\
& & - \frac{5 M_s H_s H^2}{128 \pi^2 \rho^2_s} - \frac{H^4}{32 \pi^2 \rho_s}
+ \frac{M_s^{3/2} H_s^{3/2}}{64 \pi^2 \rho^{5/2}_s} \, \sqrt{\frac{M_s H_s}{\rho_s} + H^2} \, .
\end{eqnarray}
The quantities $g^{I,{I\!I}}_0$ and $g^{I,{I\!I}}_1$ -- or, equivalently, $h^{I,{I\!I}}_0$ and $h^{I,{I\!I}}_1$ -- are kinematical Bose functions
that describe the noninteracting magnon gas. The latter are dimensionless, and in $d = d_s+1$ space-time dimensions ($d_s$=2 refers to the
spatial dimension), they are\footnote{The functions $h^{I,{I\!I}}_2$ do not show up in the free energy density, but are relevant in all other
thermodynamic quantities we derive from $z$.}
\begin{eqnarray}
\label{BoseFunctions1}
h^{I}_0(H_s, H, T) & = & \frac{4 \pi^2 {(\sigma^2 + \sigma^2_H)}^{3/2}}{3}
- 2 \sqrt{\sigma^2 + \sigma^2_H} \; Li_2(e^{2 \pi \sqrt{\sigma^2 + \sigma^2_H} }) + \frac{1}{\pi} \; Li_3(e^{2 \pi \sqrt{\sigma^2 + \sigma^2_H} }) \nonumber \\
& & + 2 \pi (\sigma^2 +\sigma^2_H) \Big\{ \log(1- e^{-2 \pi \sqrt{\sigma^2 + \sigma^2_H}}) - \log(1- e^{2 \pi  \sqrt{\sigma^2 + \sigma^2_H} }) \Big\} \, ,
\nonumber \\
h^{I}_1(H_s, H, T) & = & - \frac{1}{2 \pi} \, \log \Big( 1 - e^{- 2 \pi\sqrt{\sigma^2 + \sigma^2_H}  } \Big) \, , \nonumber \\
h^{I}_2(H_s, H, T) & = & \frac{1}{8 \pi^2 \sqrt{\sigma^2 + \sigma^2_H}  \Big( e^{2 \pi \sqrt{\sigma^2 + \sigma^2_H} } - 1 \Big)} \, ,
\end{eqnarray}
and
\begin{eqnarray}
\label{BoseFunctions2}
h^{I\!I}_0(H_s, 0, T) & = & \frac{4 \pi^2 \sigma^3}{3} + 2 \pi \sigma^2  \Big\{ \log(1- e^{-2 \pi \sigma}) - \log(1- e^{2 \pi \sigma}) \Big\}
\nonumber \\
& & - 2 \sigma \; Li_2(e^{2 \pi \sigma}) + \frac{1}{\pi} \; Li_3(e^{2 \pi \sigma}) \, , \nonumber \\
h^{I\!I}_1(H_s, 0, T) & = & - \frac{1}{2 \pi} \, \log \Big( 1 - e^{- 2 \pi \sigma} \Big) \, , \nonumber \\
h^{I\!I}_2(H_s, 0, T) & = & \frac{1}{8 \pi^2 \sigma \Big( e^{2 \pi \sigma} - 1 \Big)} \, ,
\end{eqnarray}
where $Li_2$ and $Li_3$ are polylogarithms and the two dimensionless parameters $\sigma_H$ and $\sigma$ are defined by
\begin{equation}
\label{defSigmas}
\sigma_H = \frac{H}{2 \pi T} \, , \qquad \sigma = \frac{\sqrt{M_s H_s}}{2 \pi \sqrt{\rho_s} T} \, .
\end{equation}
They capture the strength of the staggered and magnetic field relative to temperature. The connection between the functions
$h^{I,{I\!I}}_0, h^{I,{I\!I}}_1, h^{I,{I\!I}}_2$ and the dimensionful quantities $g^{I,{I\!I}}_0, g^{I,{I\!I}}_1, g^{I,{I\!I}}_2$ is provided by
\begin{eqnarray}
g^{I,{I\!I}}_0 & = & T^3 \times h^{I,{I\!I}}_0 \, , \nonumber \\
g^{I,{I\!I}}_1 & = & T \times h^{I,{I\!I}}_1 \, , \nonumber \\
g^{I,{I\!I}}_2 & = & \frac{1}{T} \times h^{I,{I\!I}}_2 \, .
\end{eqnarray}
Then, the dimensionless function $s(\sigma,\sigma_H)$ in Eq.~(\ref{freeEDtwoLoop}) collects the nontrivial part of the two-loop free energy
density coming from the sunset diagram $4d$ in Fig.~\ref{figure2}. Like the kinematical functions, $s$ depends on the staggered and
magnetic field through the two dimensionless parameters $\sigma_H$ and $\sigma$.\footnote{For the definition of $s$ and its numerical
evaluation see Ref.~\citep{Hof17}.} Finally, the quantities $e_1, e_2, k_1, k_2, k_3$ are next-to-leading order effective coupling
constants. They carry dimension of inverse energy and are small,
\begin{equation}
\label{LECs}
|e_1| \approx |e_2| \approx |k_1| \approx |k_2| \approx |k_3| \approx \frac{1}{64 \pi^3 \rho_s} \approx \frac{0.0005}{\rho_s} \, ,
\end{equation}
and only show up in the (temperature-independent) vacuum energy density.

\section{Evaluation of the Two-Point Function}
\label{TwoPointFunction}

We now derive the two-point function for the two types of magnons up to one-loop order in the effective field theory expansion. This allows
us to determine subleading corrections to the magnon dispersion relations displayed in Eq.~(\ref{disprelAFH}). In terms of these dressed
magnons we then sort out those contributions in the free energy density that can be attributed to the free Bose gas, and those that emerge
as a consequence of the magnon-magnon interaction.

\begin{figure}
\begin{center}
\includegraphics[width=12.5cm]{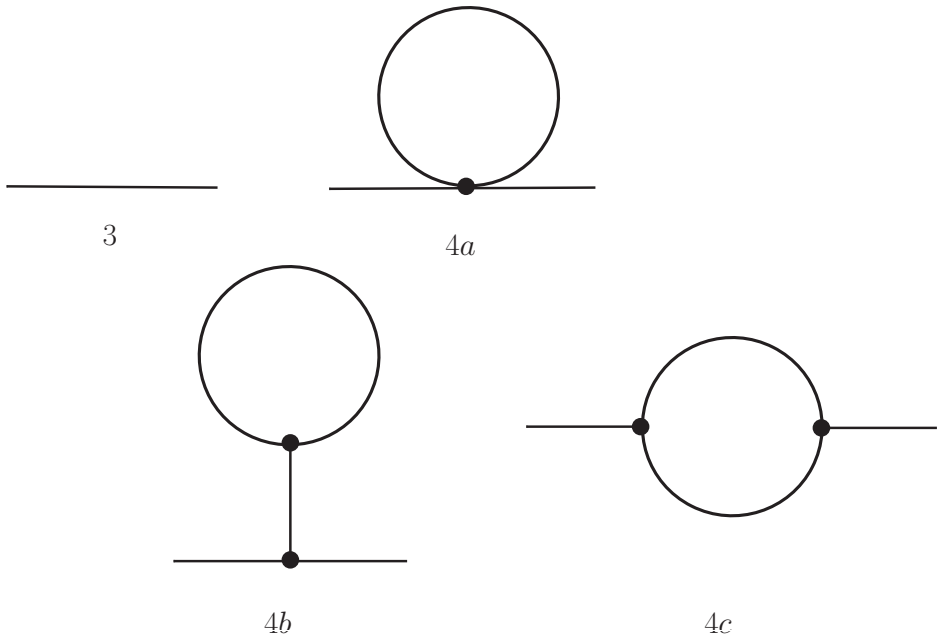}
\end{center}
\caption{Antiferromagnets in space-time dimension $d=$2+1: Feynman diagrams for the two-point function up to one-loop order in effective
field theory. Dots represent vertices from ${\cal L}^2_{eff}$.}
\label{figure1}
\end{figure}

In Fig.~\ref{figure1} we show the Feynman graphs needed to evaluate the two-point function $\tau_{I,I\!I}(x-y)$ up to the one-loop level.
The explicit expression for the leading order effective Lagrangian ${\cal L}^2_{eff}$, as well as details on the effective Lagrangian
technique, are provided in Sec. II of Ref.~\citep{Hof17}.

The dominant contribution in the two-point function $\tau_{I,I\!I}(x-y)$ comes from diagram 3,
\begin{equation}
\tau_{I,I\!I}^3(x-y) = \Delta_{I,I\!I}(x-y) = \int \!\! \frac{d^d \! k}{{(2 \pi)}^d} \, \frac{e^{ik(x-y)}}{k_0^2 + {\vec k}^2 + M^2_{I,I\!I}} \, ,
\end{equation}
where $\Delta_{I,I\!I}(x-y)$ stands for the dimensionally regularized propagators associated with the two magnons of mass $M_I$ and
$M_{I\!I}$, respectively. Subsequent corrections to the tree-level result are
\begin{eqnarray}
\label{TwoPointA}
\tau_I^{4a}(x-y) & = & \Bigg[ \Bigg( - \frac{M_s H_s}{2 \rho_s^2} - \frac{2 H^2}{\rho_s} \Bigg) \Delta_I(0) + \frac{M_s H_s}{2 \rho_s^2} 
\Delta_{I\!I}(0) \Bigg] \, \int \!\! \frac{d^d \! k}{{(2 \pi)}^d} \, \frac{e^{ik(x-y)}}{{(k_0^2 + {\vec k}^2 +  M_I^2)}^2} \, , \nonumber \\
\tau_{I\!I}^{4a}(x-y) & = & \Bigg[ \frac{M_s H_s}{2 \rho_s^2} \Big( \Delta_I(0) - \Delta_{I\!I}(0) \Big) \Bigg] \,
\int \!\! \frac{d^d \! k}{{(2 \pi)}^d} \, \frac{e^{ik(x-y)}}{{(k_0^2 + {\vec k}^2 + M_{I\!I}^2)}^2} \, , \nonumber \\
\tau_{I,I\!I}^{4b}(x-y) & = & 0 \, , \\
\tau_I^{4c}(x-y) & = & \frac{4 H^2}{\rho_s} \, \int \!\! \frac{d^d \! k}{{(2 \pi)}^d} \frac{d^d \! q}{{(2 \pi)}^d}
\frac{e^{ik(x-y)}}{{(k_0^2 + {\vec k}^2 + M_I^2)}^2} \frac{1}{q_0^2 + {\vec q}^2 + M_I^2}
\frac{{(k_0 - q_0)}^2}{{(k_0-q_0)}^2 + {({\vec k} - {\vec q})}^2 + M_{I\!I}^2} \, , \nonumber \\
\tau_{I\!I}^{4c}(x-y) & = & \frac{4 H^2}{\rho_s} \, \int \!\! \frac{d^d \! k}{{(2 \pi)}^d} \frac{d^d \! q}{{(2 \pi)}^d}
\frac{e^{ik(x-y)}}{{(k_0^2 + {\vec k}^2 + M_{I\!I}^2)}^2} \frac{1}{q_0^2 + {\vec q}^2 + M_I^2}
\frac{k_0 (k_0 - q_0)}{{(k_0-q_0)}^2 + {({\vec k} - {\vec q})}^2 + M_I^2} \, . \nonumber
\end{eqnarray}
Integration over internal momentum $q$ in $\tau_{I,I\!I}^{4c}(x-y)$ is readily done in dimensional regularization with the help of the
relations
\begin{eqnarray}
\label{IZ}
\int \!\! \frac{d^d \! q}{{(2 \pi)}^d} \frac{1}{ \Big({(p-q)}^2 + m_1^2 \Big) \Big(q^2 + m_2^2 \Big)}
& = & \frac{\Gamma(2-d/2)}{{(4 \pi)}^{d/2}} \, \int_0^1 d \alpha \ I^{d/2-2} \, , \nonumber \\
\int \!\! \frac{d^d \! q}{{(2 \pi)}^d} \frac{q_0}{ \Big({(p-q)}^2 + m_1^2 \Big) \Big(q^2 + m_2^2 \Big)}
& = & p_0 \ \frac{\Gamma(2-d/2)}{{(4 \pi)}^{d/2}} \, \int_0^1 d \alpha \ I^{d/2-2} \alpha \, , \nonumber \\
\int \!\! \frac{d^d \! q}{{(2 \pi)}^d} \frac{q_0^2}{ \Big({(p-q)}^2 + m_1^2 \Big) \Big(q^2 + m_2^2 \Big)}
& = & \frac{\Gamma(1-d/2)}{2 {(4 \pi)}^{d/2}} \, \int_0^1 d \alpha \ I^{d/2-1} \nonumber \\
& & + p_0^2 \ \frac{\Gamma(2-d/2)}{{(4 \pi)}^{d/2}} \, \int_0^1 d \alpha \ I^{d/2-2} \alpha^2 \, , \nonumber \\
& & \hspace{-6.3cm} I = \alpha (1 - \alpha) p^2 + \alpha m_1^2 + (1 - \alpha) m_2^2 \, .
\end{eqnarray}
The individual contributions, listed in Eq.~(\ref{TwoPointA}), can be incorporated into the physical two-point function $\tau_{I,I\!I}(x-y)$
through
\begin{eqnarray}
\tau_{I,I\!I}(x-y) & = & \int \!\! \frac{d^d \! k}{{(2 \pi)}^d} \, \frac{e^{ik(x-y)}}{k_0^2 + {\vec k}^2 + M^2_{I,I\!I} + X_{I,I\!I}} \\
& = & \int \!\! \frac{d^d \! k}{{(2 \pi)}^d} \, \frac{e^{ik(x-y)}}{k_0^2 + {\vec k}^2 + M^2_{I,I\!I}} \,
\Bigg\{ 1 - \frac{X_{I,I\!I}}{k_0^2 + {\vec k}^2 + M^2_{I,I\!I}} + {\cal O}(X^2/{\cal D}^2) \Bigg\} \, . \nonumber
\end{eqnarray}
Note that the dispersion law
\begin{equation}
{\cal D} = k_0^2 + {\vec k}^2 + M^2_{I,I\!I}
\end{equation}
refers to the propagator $\Delta_{I,I\!I}(x-y)$, while the quantity $X_{I,I\!I}$ involves all next-to-leading order corrections.

The physical limit $d \to 3$ does not give rise to any singularities: the $\Gamma$-functions in $\tau_{I,I\!I}^{4c}(x-y)$ [see
Eq.~(\ref{IZ})], as well as the quantities $\Delta_I(0)$ and $\Delta_{I\!I}(0)$ in $\tau_{I,I\!I}^{4a}(x-y)$, are well-defined. Accordingly,
the dispersion relations for the two magnons, in presence of mutually orthogonal magnetic and staggered fields, take the form
\begin{eqnarray}
\label{disprels}
\omega_I^2 & = & {\vec k}^2 + M^2_I + \alpha_I k_0^2 + \beta_I \, ,
\nonumber \\
\omega_{I\!I}^2 & = & {\vec k}^2 + M^2_{I\!I} + \alpha_{I\!I} k_0^2 + \beta_{I\!I} \, ,
\end{eqnarray}
with coefficients
\begin{eqnarray}
\label{dispCoeff}
\alpha_I & = & \frac{H^2 \Big(2 H^4 + 7 H^2 M^2_{I\!I} + M^3_{I\!I}(5 M_{I\!I} - 3 \sqrt{H^2 + M^2_{I\!I}})\Big)}
{8 \pi \rho_s {(M^2_{I\!I} + H^2)}^{5/2}} \nonumber \\
& & - \frac{H^2 M^2_{I\!I} (4 H^2 + M^2_{I\!I}) \log\Big(\frac{M_{I\!I} + 2 \sqrt{H^2 + M^2_{I\!I}}}{M_{I\!I}}\Big)}
{16 \pi \rho_s {(M^2_{I\!I} + H^2)}^{5/2}} \, , \nonumber \\
\beta_I & = & \frac{2 H^6 + 6 H^4 M^2_{I\!I} + H^2 M^3_{I\!I}(5 M_{I\!I} - 2 \sqrt{H^2 + M^2_{I\!I}}) + M^5_{I\!I}(M_{I\!I} - \sqrt{H^2 + M^2_{I\!I}})}
{8 \pi \rho_s {(M^2_{I\!I} + H^2)}^{3/2}} \nonumber \\
& & - \frac{H^2 M^2_{I\!I} (4 H^2 + 3 M^2_{I\!I}) \log\Big(\frac{M_{I\!I} + 2 \sqrt{H^2 + M^2_{I\!I}}}{M_{I\!I}}\Big)}
{16 \pi \rho_s {(M^2_{I\!I} + H^2)}^{3/2}} \, , \nonumber \\
\alpha_{I\!I} & = & - \frac{H^2}{4 \pi \rho_s M_{I\!I}}
\ \log \Big( \frac{M_{I\!I} - 2 \sqrt{M^2_{I\!I} + H^2}}{- M_{I\!I} - 2 \sqrt{M^2_{I\!I} + H^2}} \Big) \, , \nonumber \\
\beta_{I\!I} & = & \frac{M^2_{I\!I}(M_{I\!I} - \sqrt{M^2_{I\!I} + H^2})}{8 \pi \rho_s} \, .
\end{eqnarray}
Remember that the mass squared of magnon $I\!I$ is proportional to the staggered field,
\begin{equation}
M^2_{I\!I} = \frac{M_s H_s}{\rho_s} \, .
\end{equation}

With the above one-loop dispersion relations, the piece in the free energy density that refers to noninteracting magnons can easily be
calculated via
\begin{equation}
\label{freeEnergyBasic}
z^{free} = z_0^{free} + \frac{T}{{(2 \pi)}^2} \, \int \!\! d^2 \! k \, \ln \Big[ 1 - e^{- \omega_I({\vec k}) / T} \Big]
+ \frac{T}{{(2 \pi)}^2} \, \int \!\! d^2 \! k \, \ln \Big[ 1 - e^{- \omega_{I\!I}({\vec k}) / T} \Big] \, .
\end{equation}
The first contribution, $z_0^{free}$, is the vacuum free energy density related to noninteracting magnons. The leading term in the
dispersion law, ${\hat \omega({\vec k})}_{I,I\!I}= \sqrt{{\vec k}^2 + M^2_{I,I\!I}}$, yields the dominant contribution
$-(\mbox{$ \frac{1}{2}$} h^{I}_0 + \mbox{$ \frac{1}{2}$} h^{I\!I}_0) T^3 $ in the free energy density, Eq.~(\ref{freeEDtwoLoop}), and
corresponds to diagram 3 in Fig.~\ref{figure2}. As far as corrections $\epsilon_{I,I\!I}$ in the dispersion relations are concerned,
\begin{equation}
\omega^2_{I,I\!I} = {\vec k}^2 + M^2_{I,I\!I} + \epsilon_{I,I\!I} \, ,
\end{equation}
we expand
\begin{equation}
\exp\Big[- \frac{{\hat \omega}_{I,I\!I}}{T} \sqrt{1 + \frac{\epsilon_{I,I\!I}}{{\hat \omega}_{I,I\!I}^2}} \Big] \approx
\Bigg( 1 -\frac{\epsilon_{I,I\!I}}{2 T {\hat \omega}_{I,I\!I}}
\Bigg) \, \exp\Big[ - \frac{{\hat \omega}_{I,I\!I}}{T} \Big] \, ,
\end{equation}
and write
\begin{equation}
\ln \Big[ 1 - e^{- \sqrt{{\vec k}^2 + M^2_{I,I\!I} + \epsilon_{I,I\!I}}/ T} \Big] \approx \ln \Big[ 1 - e^{- \sqrt{{\vec k}^2 + M^2_{I,I\!I}}/ T} \Big]
+ \frac{1}{2 T} \frac{\epsilon_{I,I\!I}}{ \sqrt{{\vec k}^2 + M^2_{I,I\!I}}} \frac{1}{e^{\sqrt{{\vec k}^2 + M^2_{I,I\!I}}/T}-1} \, .
\end{equation}
Integrating over two-momentum according to Eq.~(\ref{freeEnergyBasic}), two type of contributions emerge in the free energy density,
depending on whether or not a given term in $\epsilon_{I,I\!I}$ involves $k_0^2$, that we express as\footnote{Since we restrict ourselves to
one-loop order in the evaluation of the two-point function, it is legitimate to proceed this way: higher-order loop corrections do not have
to be taken into account.}
\begin{equation}
\label{leading}
k_0^2 = - {\vec k}^2 - M^2_{I,I\!I} \, .
\end{equation}
Terms in which $\epsilon_{I,I\!I}$ is independent of $k_0$, are related to the kinematical function $h_1$, 
\begin{equation}
h_1 = \frac{1}{2 \pi T} \! \int_0^{\infty} \!\! dk \, k \, \frac{1}{\sqrt{k^2 + M^2_{I,I\!I}}} \, \frac{1}{e^{\sqrt{k^2 + M^2_{I,I\!I}}/T} - 1}
= \frac{g_1}{T} \, ,
\end{equation}
while terms where $\epsilon_{I,I\!I} \propto {\vec k}^2 $, involve the kinematical function $h_0$,
\begin{eqnarray}
h_0 & = & -\frac{1}{\pi T^2} \int_0^{\infty} \!\! dk \, k \, \ln \Big[ 1 - e^{- \sqrt{k^2 + M^2_{I,I\!I}}/T} \Big] \, , \nonumber \\
& = & \frac{1}{2 \pi T^3} \! \int_0^{\infty} \!\! dk \, k^3 \, \frac{1}{\sqrt{k^2 + M^2_{I,I\!I}}} \, \frac{1}{e^{\sqrt{k^2 + M^2_{I,I\!I}}/T} - 1}
= \frac{g_0}{T^3} \, .
\end{eqnarray}

With the above preparatory work, the piece in the free energy density that originates from the spin-wave interaction can now be extracted
as follows. Via Eq.~(\ref{freeEnergyBasic}), using the dressed magnons, we have derived the purely noninteracting part $z^{free}$.
On the other hand, in Eq.~(\ref{freeEDtwoLoop}), we have provided the full representation for the free energy density where both
interacting and noninteracting pieces are contained. By taking the difference,
\begin{equation}
z^{int} = z - z^{free} \, ,
\end{equation}
the genuine interaction piece can be isolated as
\begin{eqnarray}
\label{fedTwoLoopDRESSED}
z^{int} & = & - \frac{4 H^2 + M^2_{I\!I}}{8 \rho_s} {\Big( g^{I}_1 \Big)}^2
+ \frac{M^2_{I\!I}}{4 \rho_s} g^{I}_1 g^{I\!I}_1 - \frac{M^2_{I\!I}}{8 \rho_s} {\Big( g^{I\!I}_1 \Big)}^2
+ \frac{2}{\rho_s} \, {\tilde s} \, T^4 \nonumber \\
& & + \frac{1}{96 \pi \rho_s} \Big( {\cal I}^I_0 g^I_0 + {\cal I}^{I\!I}_0 g^{I\!I}_0 + {\cal I}^I_1 g^{I}_1 + {\cal I}^{I\!I}_1
g^{I\!I}_1 \Big) + z_0 - z_0^{free} \, .
\end{eqnarray}
The function ${\tilde s}$ is related to the function $s$ by
\begin{equation}
{\tilde s} = s -\frac{H^2}{T^4} \, R \, , 
\end{equation}
where the quantity $R$ is defined in Eq.~(B13) of Ref.~\citep{Hof17}. The respective coefficients are
\begin{eqnarray}
\label{coefficientsI}
{\cal I}^I_0 & = & \frac{2 H^2 \Big(-2 H^6 + H^4 M_{I\!I} (9M_{I\!I} - 4 M_{I}) + H^2 M^3_{I\!I} (15 M_{I\!I} + M_{I}) + 4 M^5_{I\!I} (M_{I\!I}
- M_{I}) \Big)}{M_{I}^5 {(M_{I\!I} + M_{I})}^2} \nonumber \\
& & - \frac{3 H^2 M^2_{I\!I}(4 H^2 + M^2_{I\!I})}{M^5_I} \log\Big( \frac{M_{I\!I} + 2 M_{I}}{M_{I\!I}} \Big)
\, , \nonumber \\
{\cal I}^{I\!I}_0 & = & - \frac{12 H^2}{M_{I}} + \frac{12 H^2}{M_{I\!I}} \log\Big( \frac{M_{I\!I} + 2 M_{I}}{- M_{I\!I} + 2 M_{I}} \Big)
\, , \nonumber \\
{\cal I}^I_1 & = & - \frac{4 H^2 \Big( 2 H^6 + H^4 M_{I\!I} (5 M_{I\!I} + 4 M_{I}) + H^2 M^3_{I\!I} (7 M_{I\!I} + 9 M_{I}) + 4 M^5_{I\!I}(M_{I\!I}
+ 2 M_{I})\Big)}{M^3_I {(M_{I\!I} + M_{I})}^2} \nonumber \\
& & + \frac{6 H^2 M^4_{I\!I}}{M^3_I} \log \Big( \frac{M_{I\!I} + 2 M_{I}}{M_{I\!I}} \Big)
\, , \nonumber \\
{\cal I}^{I\!I}_1 & = & - \frac{12 H^2 M^2_{I\!I}}{M_{I}} + 12 H^2 M_{I\!I} \log\Big( \frac{M_{I\!I} + 2 M_{I}}{- M_{I\!I} + 2 M_{I}} \Big) \, ,
\end{eqnarray}
with
\begin{equation}
M^2_{I} = \frac{M_s H_s}{\rho_s} + H^2 \, , \qquad M^2_{I\!I} = \frac{M_s H_s}{\rho_s} \, .
\end{equation}

Finally, the explicit expression for the free energy density associated with noninteracting magnons, calculated via
Eq.~(\ref{freeEnergyBasic}), takes the form
\begin{equation}
\label{freeDressedED}
z^{free} = - \mbox{$ \frac{1}{2}$} \Big\{ g^{I}_0 + g^{I\!I}_0 \Big\}
+ \frac{1}{32 \pi \rho_s} \Big( {\cal F}^I_0 g^I_0 + {\cal F}^{I\!I}_0 g^{I\!I}_0 + {\cal F}^I_1 g^{I}_1 + {\cal F}^{I\!I}_1 g^{I\!I}_1 \Big)
+ z_0^{free} \, ,
\end{equation}
with coefficients
\begin{eqnarray}
\label{coefficientsF}
{\cal F}^I_0 & = & - \frac{2 H^2 \Big(2 H^4 + 7 H^2 M^2_{I\!I} + M^3_{I\!I} (5 M_{I\!I} - 3 M_{I}) \Big)}{M^5_I} \nonumber \\
& & + \frac{H^2 M^2_{I\!I}(4 H^2 + M^2_{I\!I})}{M^5_I} \log\Big( \frac{M_{I\!I} + 2 M_{I}}{M_{I\!I}} \Big)
\, , \nonumber \\
{\cal F}^{I\!I}_0 & = & - \frac{4 H^2}{M_{I\!I}} \log\Big( \frac{M_{I\!I} + 2 M_{I}}{- M_{I\!I} + 2 M_{I}} \Big)
\, , \nonumber \\
{\cal F}^I_1 & = &  \frac{2M^2_{I\!I} (M^2_{I\!I} - H^2) \Big( H^2 + M_{I\!I} (M_{I\!I} - M_{I}) \Big)}{M^3_I} \nonumber \\
& & - \frac{2H^2 M^4_{I\!I}}{M^3_I} \log \Big( \frac{M_{I\!I} + 2 M_{I}}{M_{I\!I}} \Big)
\, , \nonumber \\
{\cal F}^{I\!I}_1 & = & 2 M^2_{I\!I} (M_{I\!I} - M_{I}) - 4 H^2 M_{I\!I} \log\Big( \frac{M_{I\!I} + 2 M_{I}}{- M_{I\!I} + 2 M_{I}} \Big) \, .
\end{eqnarray}

\section{Low-Temperature Series}
\label{LowTSeries}

Based on the two-loop representation for the free energy density we can now discuss the physical implications of the spin-wave interaction
in thermodynamic quantities. By "physical" we mean that we consider the {\it dressed} magnons as basic degrees of freedom in our analysis.
We first revisit pressure, order parameter and magnetization. Then we focus on the parallel staggered and uniform perpendicular
susceptibilities -- at finite as well as zero temperature.

\subsection{Scales}

The fundamental energy scale of the system is given by the exchange integral $J$. The effective field theory predictions are only valid in
the domain where temperature is low, and magnetic and staggered fields are weak, with respect to this scale. To capture the low-energy
physics of antiferromagnetic films, we introduce the dimensionless ratios,
\begin{equation}
\label{definitionRatios}
t \equiv \frac{T}{2 \pi \rho_s} \, , \qquad
m_H \equiv \frac{H}{2 \pi \rho_s} \, , \qquad
m \equiv \frac{\sqrt{M_s H_s}}{2 \pi \rho^{3/2}_s} \, .
\end{equation}
These definitions are motivated by the observation that the denominator $2 \pi \rho_s$ is of the order of $J$: loop-cluster algorithms give
the values $\rho_s = 0.18081(11) J$ and $\rho_s = 0.102(2) J$ for the square and honeycomb lattice, respectively \citep{JKNW08,JW11}. The
quantities $t, m_H, m$ hence measure temperature -- as well as magnetic and staggered field strength -- in units of the underlying energy
scale. For our effective series to be valid, they must be small. We consider the parameter domain
\begin{equation}
\label{domain}
T, \, H, \, M_{I\!I} (\propto \sqrt{H_s}) \ \lesssim 0.3 \ J \, .
\end{equation}

In contrast to antiferromagnets in three spatial dimensions, antiferromagnetic films are subjected to the Mermin-Wagner theorem
\citep{MW66}. As we pointed out on various occasions (see, e.g., Sec.~4 of Ref.~\citep{Hof14} and Sec.~V of Ref.~\citep{Hof10}), the
staggered field cannot be completely switched off: our effective representations run into trouble when one approaches the limit
$H_s \to 0$. It should be noted that no such restrictions apply to the magnetic field as it is not coupled to the order parameter: our
effective expansions remain perfectly valid in zero magnetic field.

\subsection{Pressure, Order Parameter, and Magnetization}
\label{OP}

\begin{figure}
\begin{center}
\includegraphics[width=12.5cm]{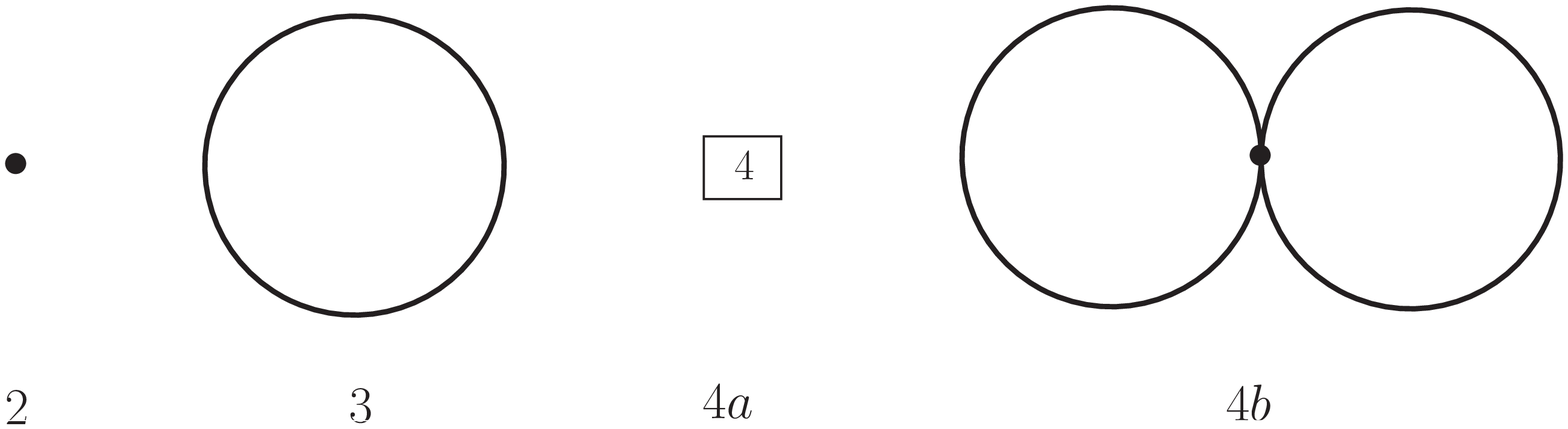}
  
\vspace{7mm}
  
\includegraphics[width=12.5cm]{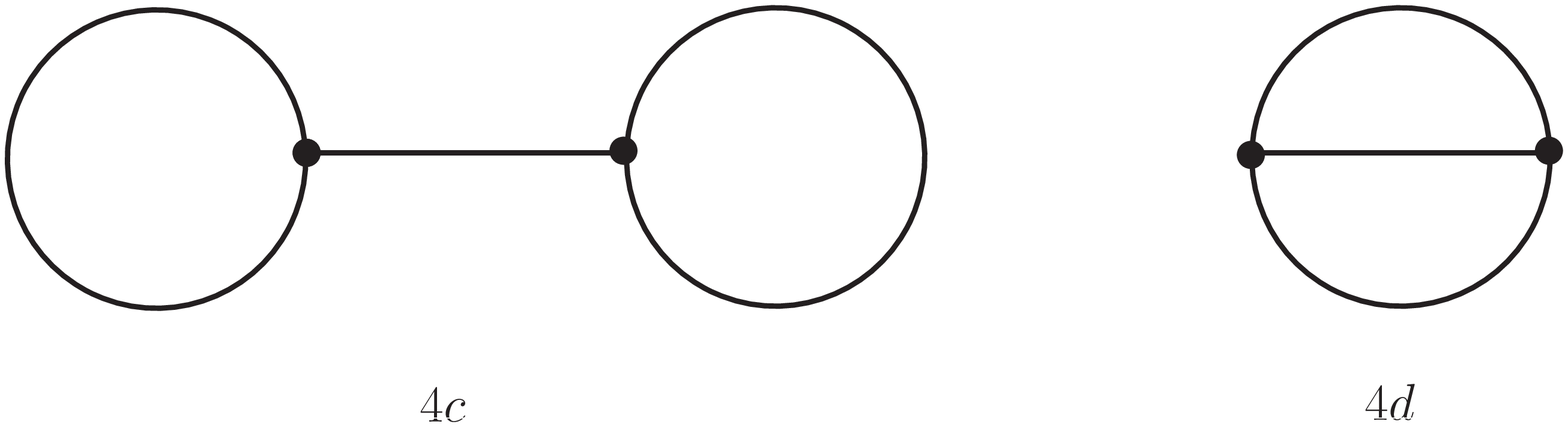}
\end{center}
\caption{Antiferromagnets in $d=$ 2+1: Feynman diagrams for the partition function or free energy density up to two-loop order $T^4$. Dots
represent vertices from ${\cal L}^2_{eff}$, while the box with the number 4 stands for the subleading piece ${\cal L}^4_{eff}$.}
\label{figure2}
\end{figure}

In the preceding paper, Ref.~\citep{Hof17}, plots have been presented to illustrate the effect of the spin-wave interaction in the
pressure, order parameter, and magnetization of antiferromagnetic films subjected to mutually orthogonal staggered and magnetic fields.
That analysis relied on a {\it diagrammatical} definition of interaction, based on the Feynman graphs for the free energy density that we
depict in Fig.~\ref{figure2}. The one-loop diagram 3 was considered as the noninteracting part, while the two-loop diagrams (where three or
four magnon lines meet and hence "interact") were classified as interaction pieces. However, as we outlined in the previous section, one
should evaluate the dispersion relations for the two types of magnons in a separate calculation that refers to zero temperature. Using
these dressed magnons, as described in the previous section, one then extracts the noninteracting piece in the free energy density -- what
is left corresponds to the genuine spin-wave interaction. The discussion below is based on this physical decomposition of the free energy
density.

\subsubsection{Pressure}

For a homogeneous system, the pressure is fixed by the temperature-dependent piece in the free energy density,
\begin{equation}
\label{defPressure}
P = - z(T) = z_0 - z \, ,
\end{equation}
where $z_0$ is the $T$=0 vacuum free energy density, and $z(T)$ refers to the purely $T$-dependent contribution. All these quantities can
be split into a piece that originates from noninteracting magnons, and a second piece that corresponds to the magnon-magnon interaction.

Using the dimensionless kinematical functions $h^{I,{I\!I}}_0, h^{I,{I\!I}}_1$, defined in Eqs.~(\ref{BoseFunctions1}) and
(\ref{BoseFunctions2}), as well as the dimensionless parameters $t, m_H, m$, defined in Eq.~(\ref{definitionRatios}), powers of
temperature in the pressure become explicit,
\begin{equation}
\label{pressureAF}
P(T,H_s,H) = {\cal P}_1 \, T^3 + \frac{{\cal P}_{\cal F} + {\cal P}_{\cal I}}{2 \pi \rho_s} \, T^4 + {\cal O}(T^5) \, .
\end{equation}
The series is dominated by the leading (one-loop) noninteracting Bose gas contribution of order $T^3$ with coefficient
\begin{equation}
{\cal P}_1 = \mbox{$ \frac{1}{2}$} \Big\{ h^{I}_0 + h^{I\!I}_0 \Big\} \, ,
\end{equation}
while the two-loop corrections are
\begin{eqnarray}
\label{pressureAFfree}
{\cal P}_{\cal F} & = & \frac{m^2_H (5 m^4 + 7 m^2_H m^2 + 2  m^4_H - 3 m^3 \sqrt{m^2 + m^2_H})}{8t {(m^2 + m^2_H)}^{5/2}} \, h^{I}_0
\nonumber \\
& & - \frac{m^2_H m^2 (m^2 + 4 m^2_H)}{16t {(m^2 + m^2_H)}^{5/2}} \, \log\Big( 1 + \frac{2 \sqrt{m^2 + m^2_H}}{m} \Big) \, h^{I}_0 \nonumber \\
& & - \frac{m^2_H}{4 m t} \, \log\Big( \frac{2 \sqrt{m^2 + m^2_H} -m}{2 \sqrt{m^2 + m^2_H} + m} \Big) \, h^{I\!I}_0
- \frac{m^2 (m^2 - m_H^2)(m^2 + m^2_H - m \sqrt{m^2 + m^2_H})}{8 t^3 {(m^2 + m^2_H)}^{3/2}} \, h^{I}_1 \nonumber \\
& & + \frac{m^2_H m^4 }{8 t^3 {(m^2 + m^2_H)}^{3/2}} \, \log\Big( 1 + \frac{2 \sqrt{m^2 + m^2_H}}{m} \Big) \, h^{I}_1 \nonumber \\
& & - \frac{m^2(m -\sqrt{m^2 + m^2_H})}{8 t^3} \, h^{I\!I}_1
- \frac{m m^2_H}{4 t^3} \, \log\Big( \frac{2\sqrt{m^2 + m^2_H}-m}{2 \sqrt{m^2 + m^2_H} + m} \Big) \, h^{I\!I}_1 \, ,
\end{eqnarray}
and
\begin{eqnarray}
\label{pressureAFinteraction}
{\cal P}_{\cal I} & = & \frac{\pi m^2}{4 t^2} \, {(h^{I}_1 - h^{I\!I}_1)}^2 + \frac{\pi m^2_H}{t^2}{(h^{I}_1)}^2
- 4 \pi \, {\tilde s}(\sigma,\sigma_H) \nonumber \\
& & - \frac{m^2_H (4 m^6 +15m^4 m^2_H +9m^4_H m^2 -2m^6_H - ( 4m^5 - m^3 m^2_H + 4 m m^4_H) \sqrt{m^2 + m^2_H})}
{24 t {(m^2 + m^2_H)}^{5/2} {{(m + \sqrt{m^2 + m^2_H})}^2}} \, h^{I}_0 \nonumber \\
& & + \frac{m^2_H m^2 (m^2 + 4 m^2_H)}{16 t {(m^2 + m^2_H)}^{5/2}} \,
\log\Big( 1 + \frac{2 \sqrt{m^2 + m^2_H}}{m} \Big) \, h^{I}_0 \nonumber \\
& & + \frac{m^2_H}{4 t \sqrt{m^2 + m^2_H}} \, h^{I\!I}_0
+ \frac{m^2_H}{4 m t} \, \log\Big( \frac{2 \sqrt{m^2 + m^2_H} -m}{2 \sqrt{m^2 + m^2_H} + m} \Big) \, h^{I\!I}_0 \nonumber \\
& & + \frac{m^2_H (4 m^6 + 7m^4 m^2_H + 5m^4_H m^2 + 2m^6_H + ( 8m^5 + 9m^3 m^2_H + 4 m m^4_H) \sqrt{m^2 + m^2_H})}
{12 t^3 {(m^2 + m^2_H)}^{3/2}{(m + \sqrt{m^2 + m^2_H})}^2} \, h^{I}_1
\nonumber \\
& & - \frac{m^2_H m^4}{8 t^3 {(m^2 + m^2_H)}^{3/2}} \,
\log\Big( 1 + \frac{2 \sqrt{m^2 + m^2_H}}{m} \Big) \, h^{I}_1 \nonumber \\
& & + \frac{m^2_H m^2}{4 t^3 \sqrt{m^2 + m^2_H}} \, h^{I\!I}_1
+ \frac{m m^2_H}{4 t^3} \, \log\Big( \frac{2\sqrt{m^2 + m^2_H}-m}{2 \sqrt{m^2 + m^2_H} + m} \Big) \, h^{I\!I}_1 \, .
\end{eqnarray}
The contribution ${\cal P}_{\cal I} T^4$ represents the genuine spin-wave interaction. In the limit $H \to 0$, the terms ${\cal P}_{\cal F}$
and ${\cal P}_{\cal I}$ vanish identically: in the absence of the magnetic field, spin waves only interact beyond two-loop order.

\begin{figure}
\begin{center}
\hbox{
\includegraphics[width=8.0cm]{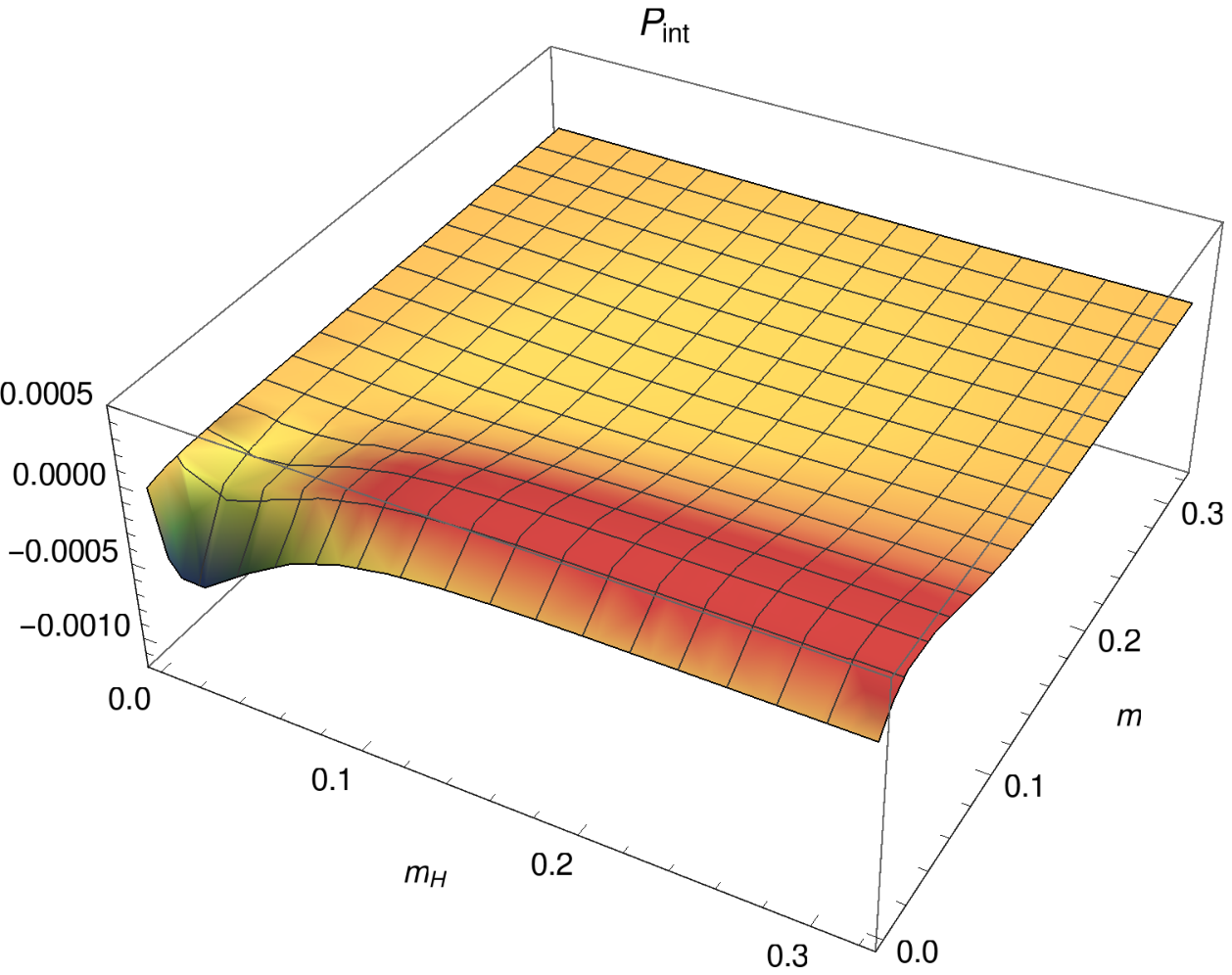} 
\includegraphics[width=8.0cm]{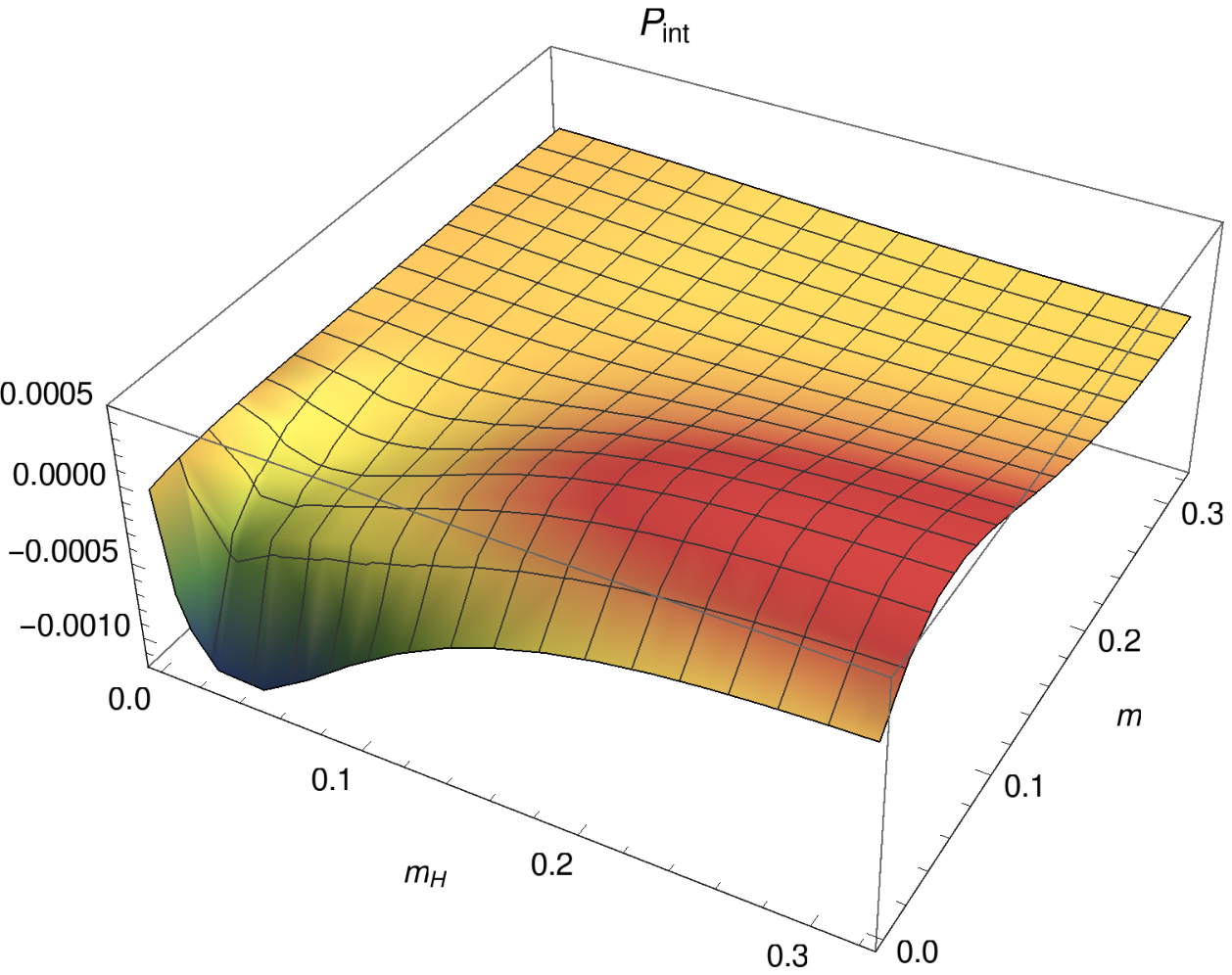}}
\vspace{7mm}
\hbox{
\includegraphics[width=8.0cm]{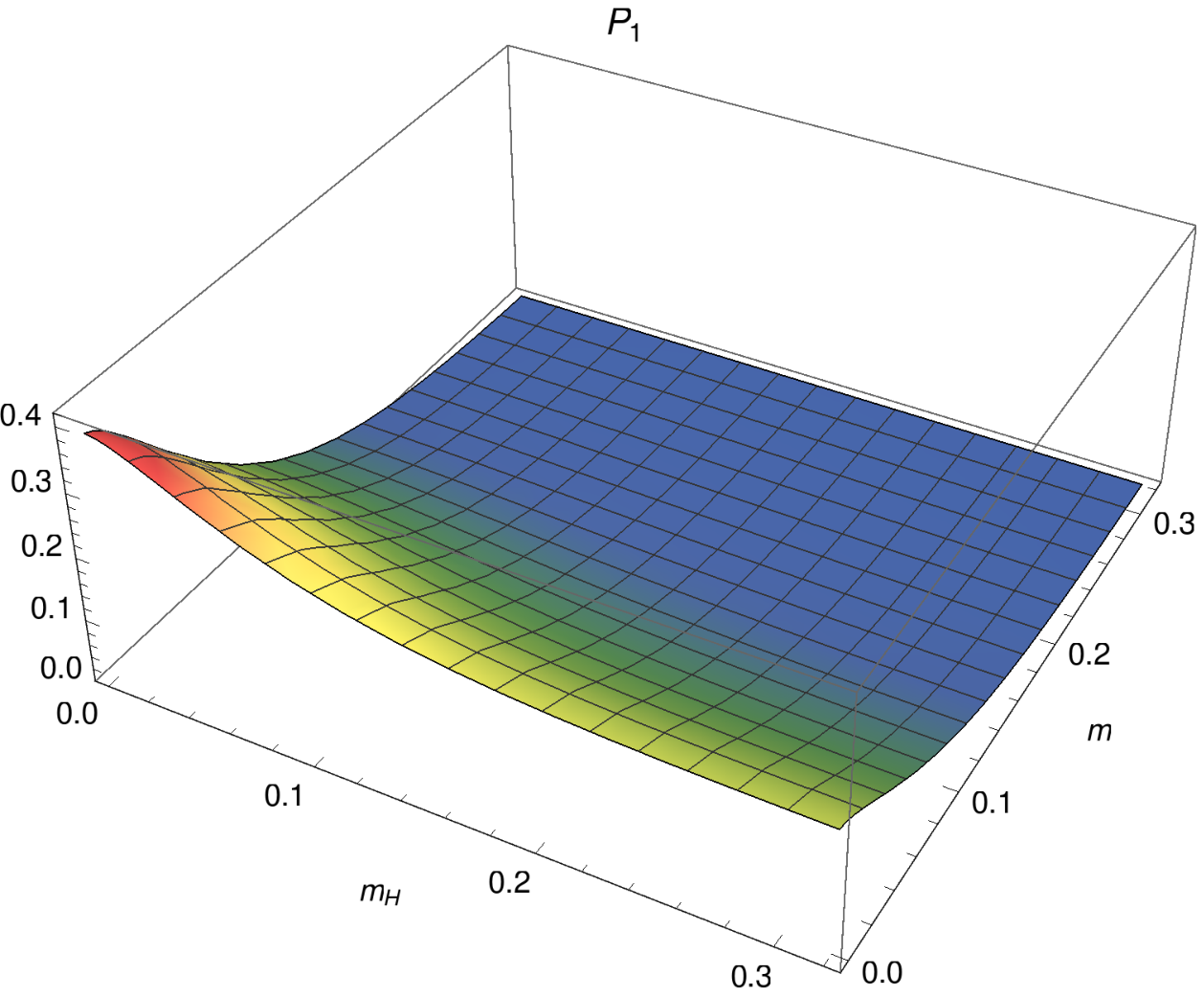}
\includegraphics[width=8.0cm]{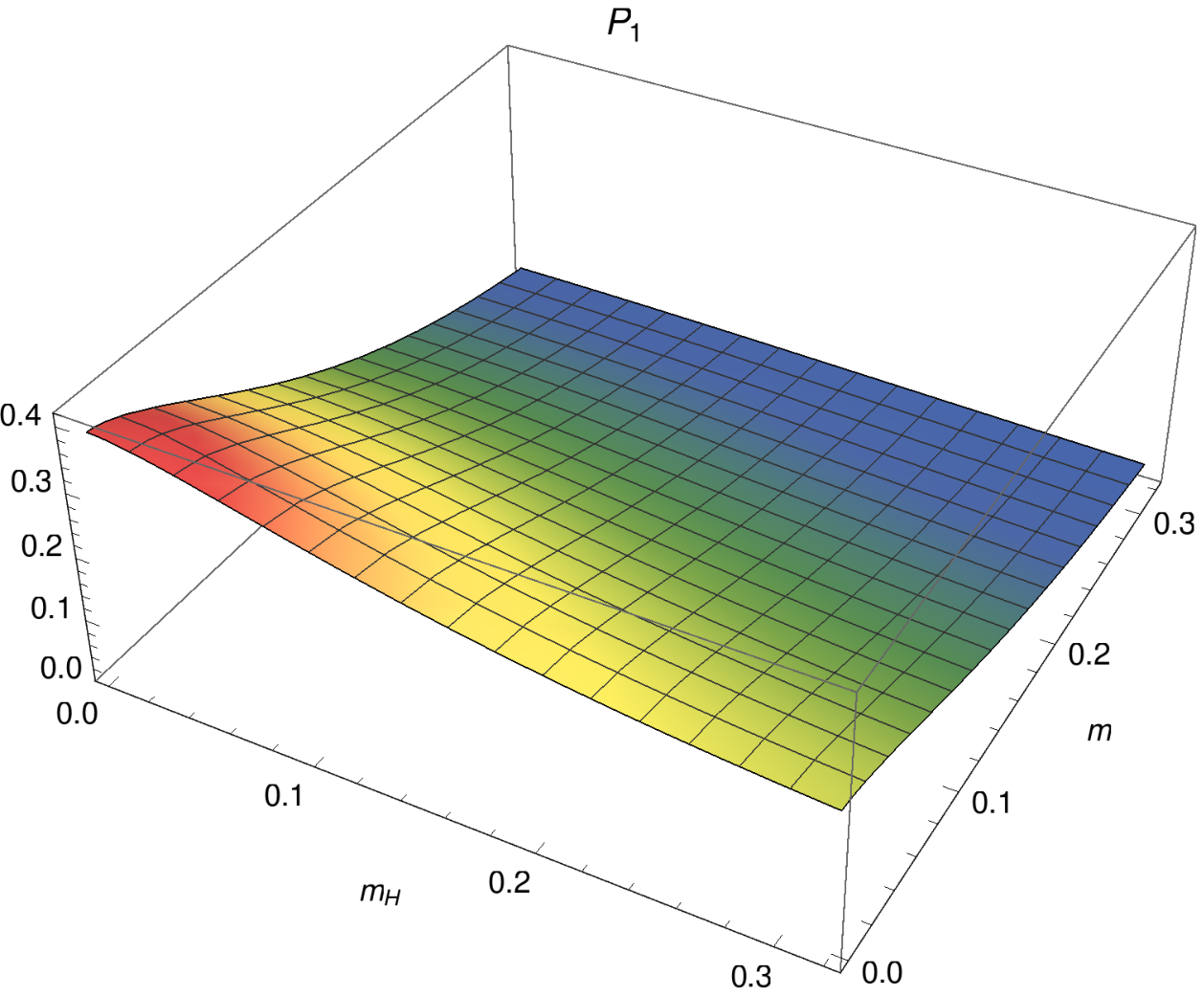}}
\end{center}
\caption{[Color online] Upper panel: Genuine spin-wave interaction manifesting itself in the pressure -- measured by
$P_{int} = {\cal P}_{\cal I} \, t$ --  of antiferromagnetic films in mutually orthogonal magnetic and staggered fields at the temperatures
$T/2 \pi \rho_s = 0.05$ (left) and $T/2 \pi \rho_s = 0.1$ (right). Lower panel: Leading contribution ${\cal P}_1$ in the pressure for the
same two temperatures $T/2 \pi \rho_s = 0.05$ (left) and $T/2 \pi \rho_s = 0.1$ (right).}
\label{figure3}
\end{figure}

To assess the effect of the magnon-magnon interaction, we consider the dimensionless ratio\footnote{See Eq.~(\ref{pressureAF}).}
\begin{equation}
\label{intRatioP}
P_{int} = \frac{{\cal P}_{\cal I} \; T^4}{2 \pi  \rho_s} \, \frac{1}{T^3} = {\cal P}_{\cal I} \, t \, ,
\end{equation}
that measures sign and strength of the interaction. In Fig.~\ref{figure3}, on the upper panel, we plot $P_{int}$ for the two temperatures
$t = T/2 \pi \rho_s = 0.05$ and $t = T/2 \pi \rho_s = 0.1$. On the lower panel we display the dominant contribution in the pressure --
${\cal P}_1$ -- that refers to noninteracting magnons. One notices that the spin-wave interaction in the pressure is very small, less than
one percent relative to the free Bose gas contribution. Overall, the genuine spin-wave interaction is repulsive in most of parameter space.
Interestingly, in weaker staggered and magnetic fields, the interaction becomes attractive. However, one should not forget the fact that we
are describing very subtle effects.

\subsubsection{Staggered Magnetization}

\begin{figure}
\begin{center}
~~~~\hbox{
\includegraphics[width=7.5cm]{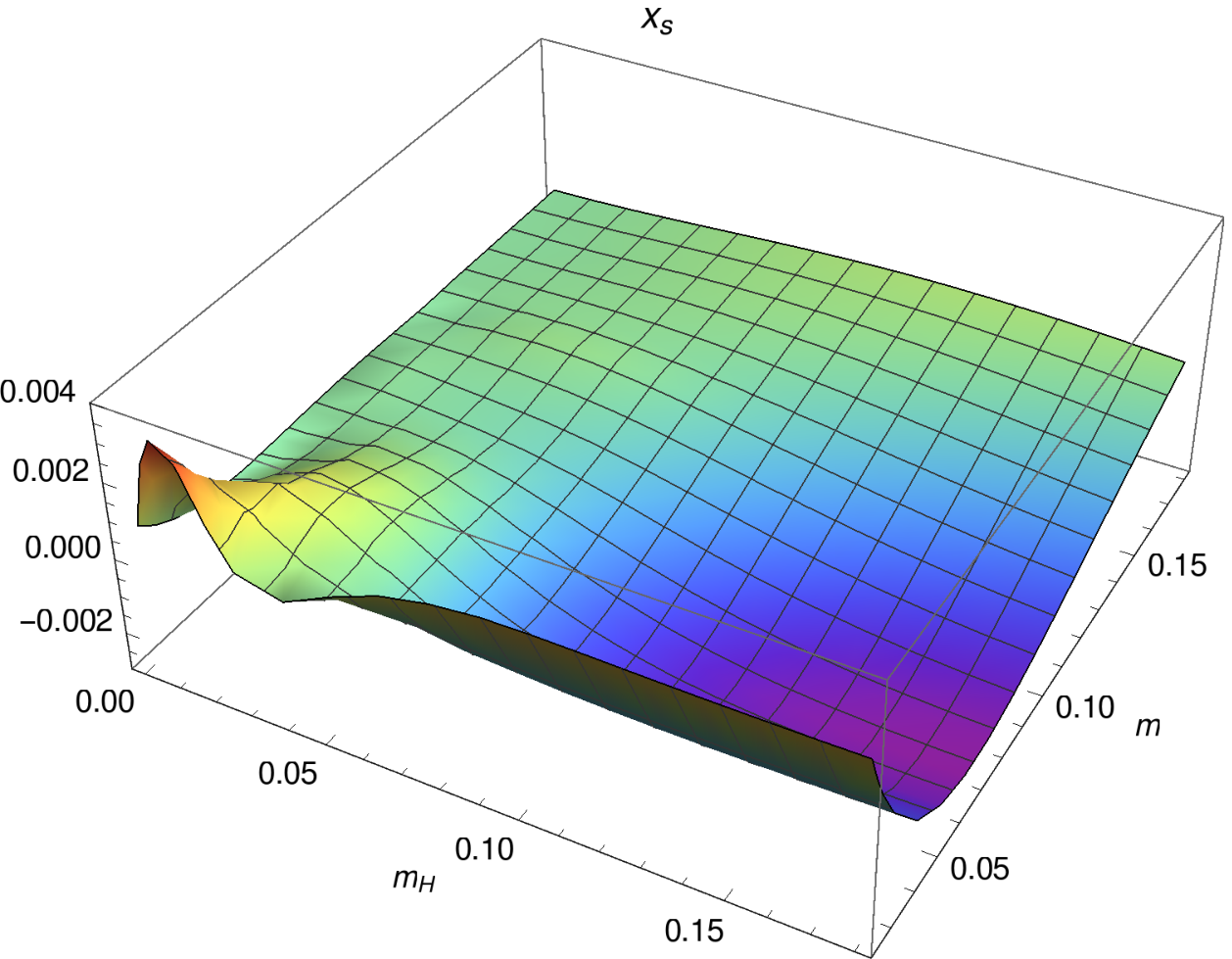}
\includegraphics[width=7.5cm]{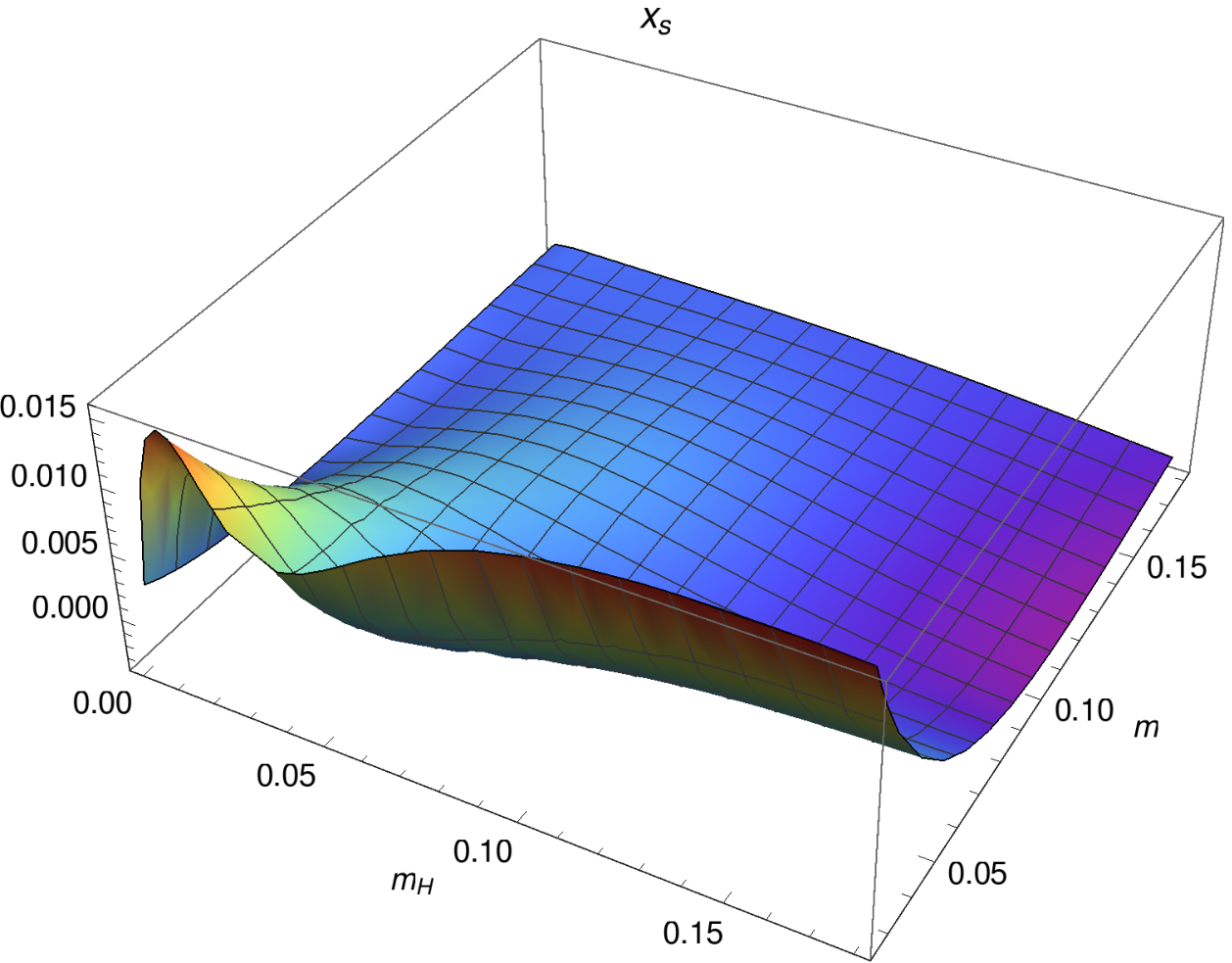}}
\end{center}
\caption{[Color online] Genuine spin-wave interaction manifesting itself in the staggered magnetization -- measured by $x_s(T,H_s,H)$ -- of
antiferromagnetic films in mutually orthogonal magnetic and staggered fields at the temperatures $T/2 \pi \rho_s = 0.05$ (left) and
$T/2 \pi \rho_s = 0.1$ (right).}
\label{figure4}
\end{figure}

We proceed with the staggered magnetization,
\begin{equation}
M_s(T,H_s,H) = - \frac{\partial z(T,H_s,H)}{\partial H_s} \, ,
\end{equation}
where we obtain
\begin{eqnarray}
\label{OPAF}
& & M_s(T,H_s,H) = M_s(0,H_s,H) + {\cal M}_s^{[1]} T + ( {\cal M}_s^{\cal F} + {\cal M}_s^{\cal I} ) T^2 + {\cal O}(T^3) \, , \nonumber \\
& & {\cal M}_s^{[1]}  = -\frac{M_s}{2 \rho_s} \, \Big( h^{I}_1 + h^{I\!I}_1 \Big) \, .
\end{eqnarray}
The interaction sets in at order $T^2$ and is represented by the coefficient ${\cal M}_s^{\cal I} $. Accordingly, we define the parameter
$x_s$,
\begin{equation}
x_s(T,H_s,H) = \frac{ {\cal M}_s^{\cal I} \; T^2}{| {\cal M}_s^{[1]} \; T |} \, ,
\end{equation}
that measures the effect of the interaction with respect to the leading free Bose gas contribution. As we show in Fig.~\ref{figure4}, the
impact of the genuine spin-wave interaction in the staggered magnetization is weak. The parameter $x_s$ may take negative or positive
values. In particular, in weaker staggered fields it becomes positive. $x_s>0$ means that if temperature is increased from $T$=0 to finite
$T$, the staggered magnetization grows (while keeping the field strengths $H$ and $H_s$ fixed) as a consequence of the magnon-magnon
interaction. Of course, the behavior of the staggered magnetization is dominated by the (negative) one-loop contribution ${\cal M}_s^{[1]}$
that causes a decrease of the staggered magnetization at finite temperature for any of the values $H$ and $H_s$ we consider.\footnote{See
Eq.~(\ref{OPAF}) and Fig.~\ref{figure6}.}

\subsubsection{Magnetization}

\begin{figure}
\begin{center}
~~~~\hbox{
\includegraphics[width=7.5cm]{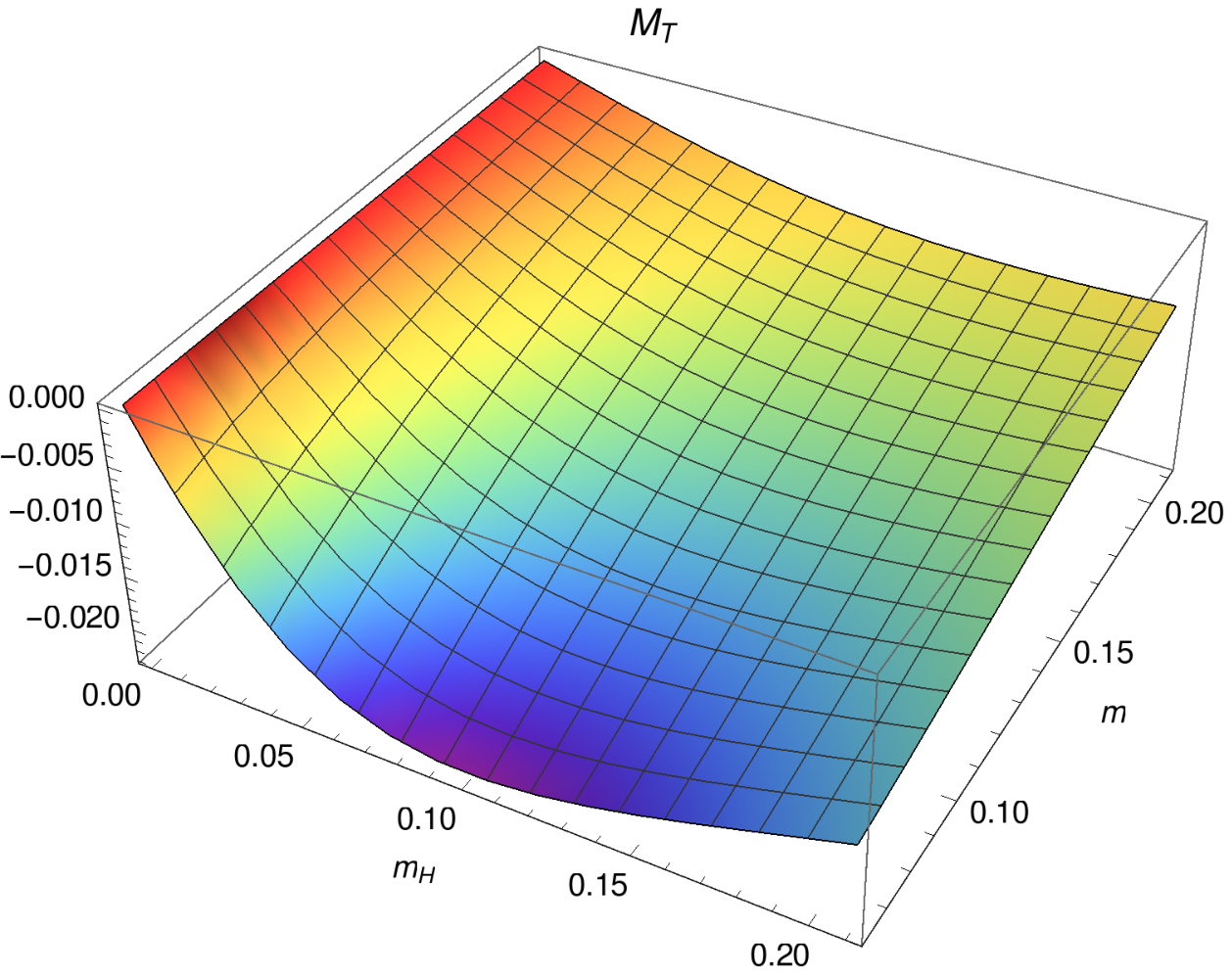}
\includegraphics[width=7.5cm]{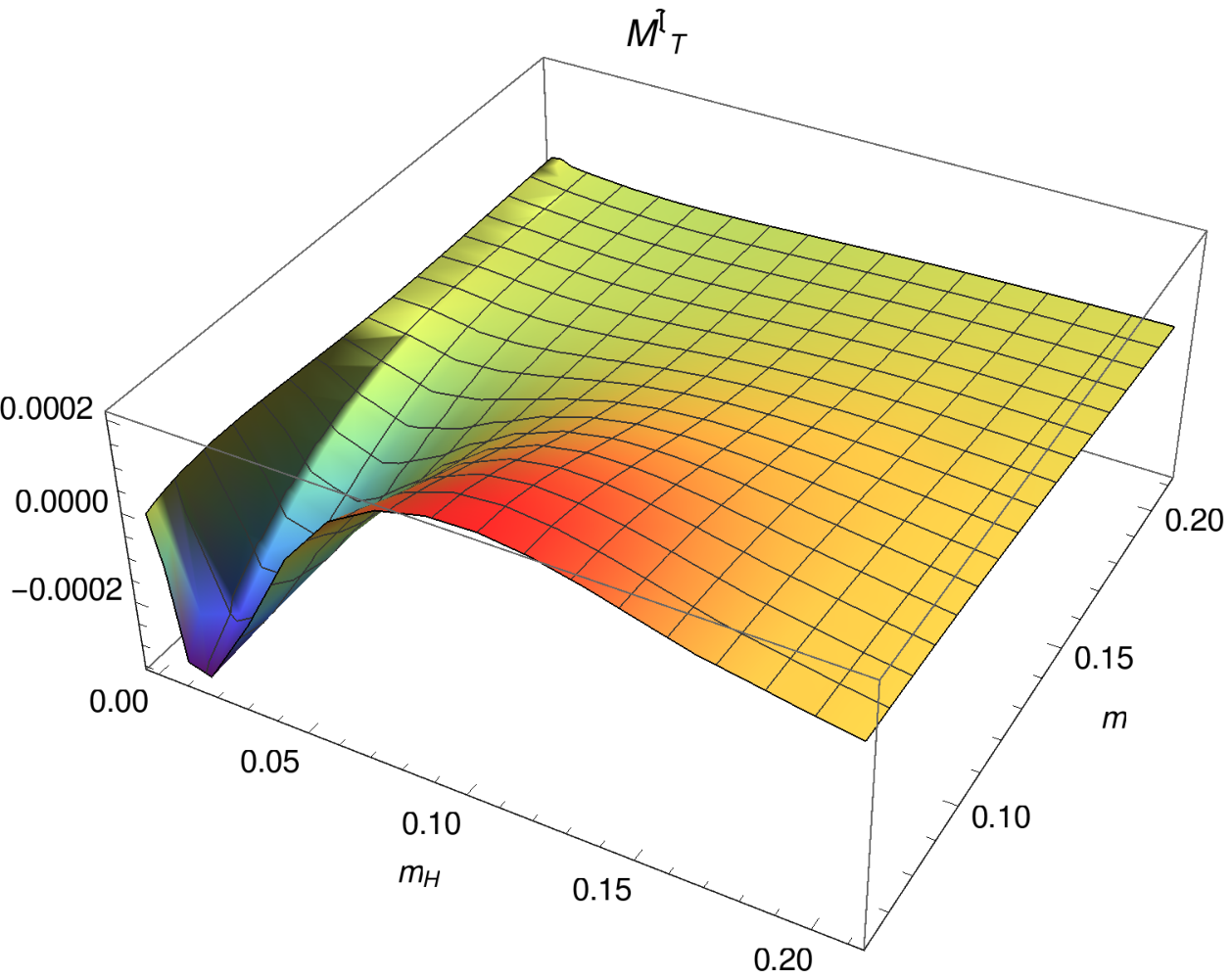}}
\end{center}
\caption{[Color online] Antiferromagnetic films in mutually orthogonal magnetic and staggered fields at $T/2 \pi \rho_s = 0.1$. Left:
Full temperature-dependent part of the magnetization $M_T(T,H_s,H)$. Right: Effect of the genuine spin-wave interaction in the
magnetization measured by $M^{\cal I}_T(T,H_s,H)$.}
\label{figure5}
\end{figure}

For the low-temperature series of the magnetization,
\begin{equation}
M(T,H_s,H) = - \frac{\partial z(T,H_s,H)}{\partial H} \, ,
\end{equation}
we get
\begin{eqnarray}
\label{magnetizationAF}
& & M(T,H_s,H) = M(0,H_s,H) + {\cal M}^{[1]} T + ( {\cal M}^{\cal F} + {\cal M}^{\cal I} ) T^2 + {\cal O}(T^3) \, , \nonumber \\
& & {\cal M}^{[1]}(T,H_s,H) = - H h^{I}_1 \, .
\end{eqnarray}
The full temperature-dependent part in the magnetization,
\begin{equation}
\label{SigT}
M_T(T,H_s,H) = \frac{{\cal M}^{[1]} \; T + ( {\cal M}^{\cal F} + {\cal M}^{\cal I} ) \; T^2 }{\rho^2_s} \, ,
\end{equation}
always takes negative values, as we depict on the left-hand side of Fig.~\ref{figure5} for the temperature $T/2 \pi \rho_s = 0.1$. If
temperature is increased from $T$=0 to finite $T$, the magnetization drops (while keeping the field strengths $H$ and $H_s$ fixed), as one
expects. However, subtle effects show up when one focuses on the spin-wave interaction,
\begin{equation}
M^{\cal I}_T(T,H_s,H) = \frac{{\cal M}^{\cal I} T^2}{\rho^2_s} \, ,
\end{equation}
as we illustrate on the right-hand side of Fig.~\ref{figure5} for the same temperature $T/2 \pi \rho_s = 0.1$. Notice that the interaction
contribution is about two orders of magnitude smaller than $M_T(T,H_s,H)$ in the parameter region we have displayed. Remarkably,
$M^{\cal I}_T(T,H_s,H)$ may acquire negative or positive values, depending on the field strengths $H$ and $H_s$. Positive values are somehow
counterintuitive as they mean that if temperature is raised from $T$=0 to finite $T$ (while keeping the field strengths fixed), the
magnetization increases as a consequence of the spin-wave interaction.

\subsection{Parallel Staggered and Uniform Perpendicular Susceptibility}

The parallel staggered and uniform perpendicular susceptibilities\footnote{The uniform perpendicular susceptibility is also referred to as
uniform transverse susceptibility. It measures the response of the system with respect to an external magnetic field that is oriented
perpendicular to the staggered magnetization (or staggered field).} are defined by 
\begin{equation}
\label{defStaggeredSusceptibity}
\chi^s_{\parallel}(T,H_s,H) = \frac{\partial M_s(T,H_s,H)}{\partial H_s} \, ,
\end{equation}
and
\begin{equation}
\label{defUniformSusceptibity}
\chi_{\perp}(T,H_s,H) = \frac{\partial M(T,H_s,H)}{\partial H} \, ,
\end{equation}
respectively.

\subsubsection{Zero Temperature}

Let us first consider the situation at zero temperature, where the two quantities take the form
\begin{eqnarray}
\frac{32 \pi^2 \rho_s^3}{M^2_s} \, \chi^s_{\parallel}(0,H_s,H) & = & \frac{1}{m} + \frac{1}{\sqrt{m^2+m_H^2}} + 1 + 64 \pi^2 \rho_s (k_2 + k_3)
\nonumber \\
& & - \frac{8 m^4 + 12 m^2 m_H^2 + 3 m_H^4}{8 m {(m^2+m_H^2)}^{3/2}} \, ,
\end{eqnarray}
and
\begin{eqnarray}
\frac{\chi_{\perp}(0,H_s,H)}{\rho_s} & = & 1 + \frac{m^2 + 2 m_H^2}{2 \sqrt{m^2+m_H^2}}
+ 8 \pi^2 \rho_s \Big\{k_1 m^2 + 6(e_1 + e_2) m_H^2 \Big\} \\
& & + \frac{- m^5 + 5 m^4 \sqrt{m^2+m_H^2} + 29 m^2 m_H^2 \sqrt{m^2+m_H^2} + 24 m_H^4 \sqrt{m^2+m_H^2}}{16{(m^2+m_H^2)}^{3/2}} \, . \nonumber
\end{eqnarray}
Recall that the dimensionless parameters $m$ and $m_H$,
\begin{equation}
m \equiv \frac{\sqrt{M_s H_s}}{2 \pi \rho_s^{3/2}} \, , \qquad
m_H \equiv \frac{H}{2 \pi \rho_s} \, ,
\end{equation}
measure the strength of the staggered and magnetic field. Note that, at $T$=0, next-to-leading order low-energy couplings -- $e_1, e_2,
k_1, k_2$, and $k_3$ -- show up in the zero-temperature susceptibilities.\footnote{We stress that finite-temperature expressions are free
of such next-to-leading order low-energy couplings, meaning that effective field theory predictions for observables at $T \neq 0$ are
parameter-free.} According to Eq.~(\ref{LECs}) their numerical values are small, and they only matter at subleading orders.

In zero magnetic field, the parallel staggered susceptibility reduces to
\begin{equation}
\chi^s_{\parallel}(0,H_s,0) = \frac{M_s^{3/2}}{8 \pi \rho_s^{3/2}} \, \frac{1}{\sqrt{H_s}} + \frac{2 M_s^2}{\rho_s^2} \, (k_2 + k_3)
+ {\cal O}(\sqrt{H_s}) \, ,
\end{equation}
i.e., it diverges like $\propto 1/\sqrt{H_s}$. We point out that this effective theory prediction is perfectly valid in the limit
$H_s \to 0$, since we are at zero temperature where the Mermin-Wagner theorem does not apply. Hence the divergence of $\chi^s_{\parallel}$
is not an artifact of the effective calculation, but is physical.

As far as the uniform perpendicular susceptibility is concerned, both limits $H_s \to 0$ and $H \to 0$ are well-defined at zero
temperature,
\begin{eqnarray}
\chi_{\perp}(0,0,H) & = & \rho_s + \frac{H}{2 \pi} + \Big\{ 12(e_1 + e_2) + \frac{3}{8 \pi^2 \rho_s} \Big\} H^2 + {\cal O}(H^3) \, ,
\nonumber \\
\chi_{\perp}(0,H_s,0) & = & \rho_s + \frac{\sqrt{M_s}}{4 \pi \sqrt{\rho_s}} \, \sqrt{H_s}
+ \Big\{ \frac{2 k_1 M_s}{\rho_s} + \frac{M_s}{16 \pi^2 \rho_s^2} \Big\} \, H_s  + {\cal O}(H_s^{3/2}) \, .
\end{eqnarray}
If both fields are switched off, we obtain the simple result
\begin{equation}
\label{susHelModulus}
\chi_{\perp}(0,0,0) = \rho_s \, .
\end{equation}
This connection between uniform perpendicular susceptibility, helicity modulus and spin-wave velocity,\footnote{We temporarily restore
dimensions in order to compare with the literature. In the effective field theory formalism, the spin-wave velocity is dimensionless and
set to one.}
\begin{equation}
\label{connectionChiRho}
\chi_{\perp}(0,0,0) = \frac{\rho_s}{v^2} \, ,
\end{equation}
has been derived a long time ago within hydrodynamic theory in the pioneering paper by Halperin and Hohenberg \citep{HH69}. Over the years,
$\chi_{\perp}(0,0,0)$ has been calculated by various authors with different methods such as Schwinger boson mean-field, series expansions
around the Ising limit, and Green's-function Monte Carlo methods \citep{Sin89,AA88b,WOH91,Run92,Iga92,CW93,HWO94}. On the other hand,
$\rho_s$ as well as $v$, have been determined to high precision by very efficient loop-cluster algorithms. The most recent data for the
square lattice are \citep{JW11}
\begin{equation}
\rho_s = 0.18081(11) \; J \, , \qquad v = 1.6586(3) \; J a , \qquad (S=\frac{1}{2}) \, .
\end{equation}
This then leads to the very accurate result for the uniform perpendicular susceptibility,
\begin{equation}
\chi_{\perp}(0,0,0) = \frac{\rho_s}{v^2} = 0.06573(3) \; 1/Ja^2 \, , \qquad (S=\frac{1}{2}) \, ,
\end{equation}
consistent with the directly calculated values for $\chi_{\perp}$ in Refs.~\citep{AA88b,Sin89,WOH91,Run92,Iga92,HWO94}.

Loop-cluster results are also available for the honeycomb lattice \citep{JKNW08},
\begin{equation}
\rho_s = 0.102(2) \; J \, , \qquad v = 1.297(16) \; J a \, , \qquad (S=\frac{1}{2}) \, ,
\end{equation}
where we obtain the value
\begin{equation}
\chi_{\perp}(0,0,0) = \frac{\rho_s}{v^2} = 0.061(1) \; 1/Ja^2 \, , \qquad (S=\frac{1}{2}) \, .
\end{equation}
We are not aware of any direct calculations of $\chi_{\perp}(0,0,0)$ for the honeycomb geometry.

\subsubsection{Finite Temperature}

Let us now turn to finite temperature where the parallel staggered and uniform perpendicular susceptibilities have been analyzed in many
articles \citep{Gho73,AA88b,CHN88,CHN89,GJN89,Tak89a,Tak89b,OY89,Tak90a,Tak90b,OY90,MD91,SR91,Man91,Bar91,HN93,AS93,CSY94,CTVV97,KT98,HTK98,
SS99,KKM00,CRTVV00,SS01,SLZ09}. The various methods comprise modified spin-wave theory, high-temperature expansion, finite-size scaling
analysis, Monte Carlo simulations, Green's function methods, and yet other approaches. Unfortunately, a direct comparison with our effective
field theory results is not possible, simply because the above-mentioned articles refer to the finite-temperature susceptibilities in
{\bf zero staggered field} -- this, of course, is the physically most relevant case. Our effective analysis, however, refers to the domain
where the staggered field -- although weak compared to the underlying scale fixed by the exchange coupling $J$ -- cannot be switched off.
It is not consistent to take the limit $H_s \to 0$ in our effective calculation, as we have mentioned earlier. As for the pressure, order
parameter, and magnetization, in this section we are interested in how the susceptibilities behave in nonzero staggered and magnetic fields
and how the spin-wave interaction emerges at finite temperature. Apart from being completely systematic, our two-loop results hence appear
to be entirely new.

In effective field theory, the low-temperature series for the parallel staggered and the uniform perpendicular susceptibility take the form
\begin{eqnarray}
& & \chi^s_{\parallel}(T,H_s,H) = \chi^s_{\parallel}(0,H_s,H) + \frac{\chi^{[1]}_s}{T} + \chi^{\cal F}_s + \chi^{\cal I}_s + {\cal O}(T) \, ,
\nonumber \\
& & \chi_{\perp}(T,H_s,H) = \chi_{\perp}(0,H_s,H) + \chi^{[1]} \; T  + ( \chi^{\cal F} + \chi^{\cal I}) T^2 + {\cal O}(T^3) \, , \nonumber \\
& & \qquad \chi^{[1]}_s(T,H_s,H) = \frac{M^2_s}{2 \rho_s^2} \, \Big( h^{I}_2 + h^{I\!I}_2 \Big) \, , \nonumber \\
& & \qquad \chi^{[1]}(T,H_s,H) = \frac{2 H^2}{T^2} \, h^I_2 - h^I_1 \, .
\end{eqnarray}
The spin-wave interaction emerges at two-loop order in both susceptibilities: the respective temperature powers are $T^0$ ($T^2$) in the
parallel staggered (uniform perpendicular) susceptibility. It should be emphasized that the corresponding two-loop coefficients
$\chi^{\cal I}_s,\chi^{\cal I}$ -- much like $\chi^{\cal F}_s,\chi^{\cal F}$ and $\chi^{[1]}_s$ and $\chi^{[1]}$ -- depend in a nontrivial way on the
dimensionless ratios $t, m_H, m$ defined in Eq.~(\ref{definitionRatios}). The explicit expressions are rather involved and will not be
listed here. Note that they can be obtained trivially from the two-loop representation of the pressure provided in
Eqs.~(\ref{pressureAF}), (\ref{pressureAFfree}) and (\ref{pressureAFinteraction}). On the other hand, the quantities
$\chi^s_{\parallel}(0,H_s,H)$ and $\chi_{\perp}(0,H_s,H)$ that we discussed previously, only involve $m$ and $m_H$, but are $T$-independent.

\begin{figure}
\begin{center}
\includegraphics[width=10.5cm]{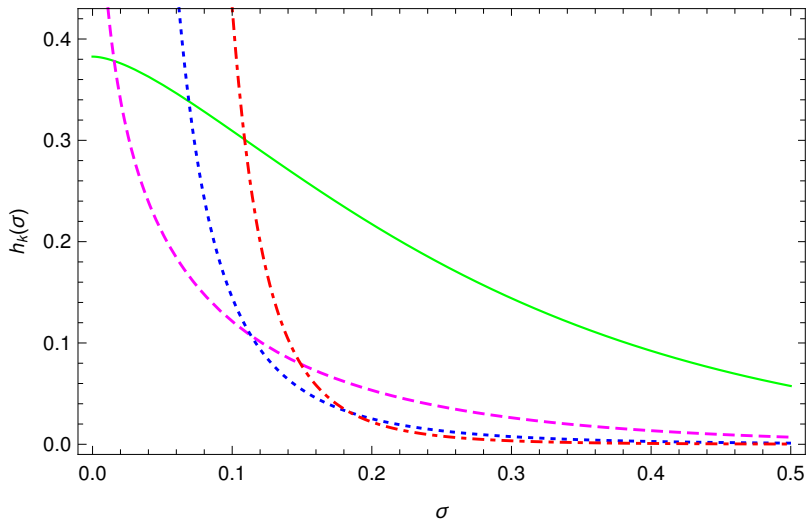}
\end{center}
\caption{[Color online] The kinematical functions $h^I_0(\sigma,0)$ (continuous), $h^I_1(\sigma,0)$ (dashed), $h^I_2(\sigma,0)$ (dotted),
and $h^I_3(\sigma,0)$ (dashed-dotted), as a function of the parameter $\sigma$. Whereas $h^I_0(\sigma,0)$ tends to a finite value in the
limit $\sigma \to 0$, the other kinematical functions diverge.}
\label{figure6}
\end{figure}

The one-loop contributions $\chi^{[1]}_s$ and $\chi^{[1]}$ are proportional to the kinematical functions
$h^I_1(\sigma, \sigma_H),h^I_2(\sigma, \sigma_H)$ and $h^{I\!I}_2(\sigma)$.\footnote{The parameters $\sigma$ and $\sigma_H$ are defined in
Eq.~(\ref{defSigmas}).} Notice that the functions $h^I_k(\sigma, \sigma_H)$ that refer to magnon $I$, are related to the functions
$h^{I\!I}_k(\sigma)$ that refer to magnon $I\!I$, via\footnote{See Eqs.~(\ref{BoseFunctions1}) and (\ref{BoseFunctions2}).}
\begin{equation}
h^{I\!I}_k(\sigma) = h^I_k(\sigma, \sigma_H=0) \, , \qquad k=0,1,2,3 \, .
\end{equation}
In Fig.~\ref{figure6} we depict all four functions $h^I_k(\sigma,0), \; k=0,1,2,3$. While $h^I_0(\sigma, 0)$ tends to the finite value
$\zeta(3)/\pi$ in the double limit $\{ H_s, H \} \to 0$, the other functions diverge when both fields are switched off. Accordingly, in the
absence of the magnetic field, the leading terms $\chi^{[1]}_s$ and $\chi^{[1]}$ become singular in the limit $H_s \to 0$. One should keep in
mind, however, that this limit is not legitimate at finite temperature in our effective formalism. Unlike at $T$=0, the divergence of the
parallel staggered (and uniform perpendicular) susceptibility at $T \neq 0$ is an artifact of our calculation. Therefore one has to be
careful not to leave the parameter domain beyond which our low-temperature series become invalid. By inspecting Figs. 2 and 3 of
Ref.~\citep{Hof16a}, the interested reader may verify that in all plots shown in the present work, we are well within the allowed parameter
domain.

\begin{figure}
\begin{center}
~~~~\hbox{
\includegraphics[width=7.5cm]{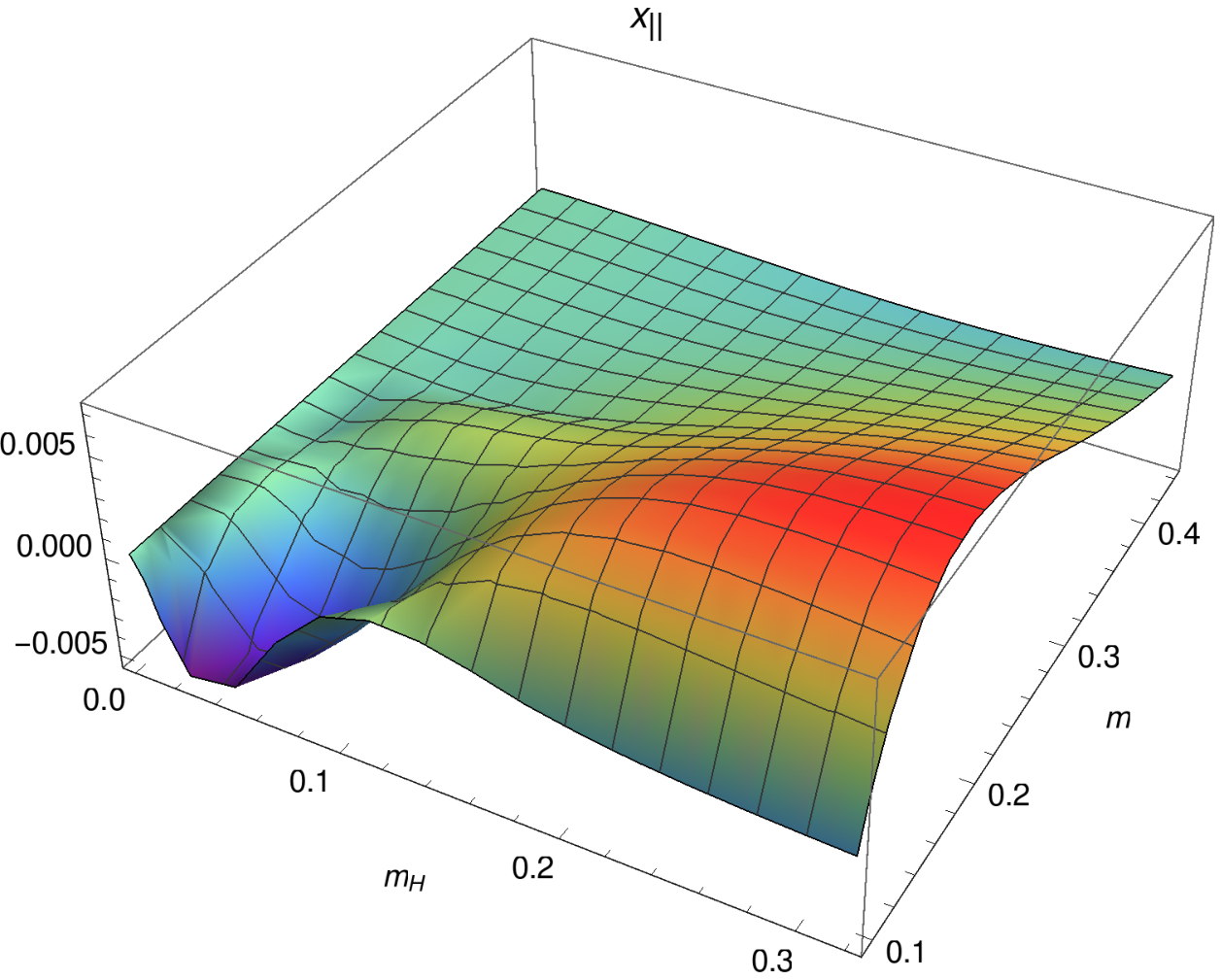}
\includegraphics[width=7.5cm]{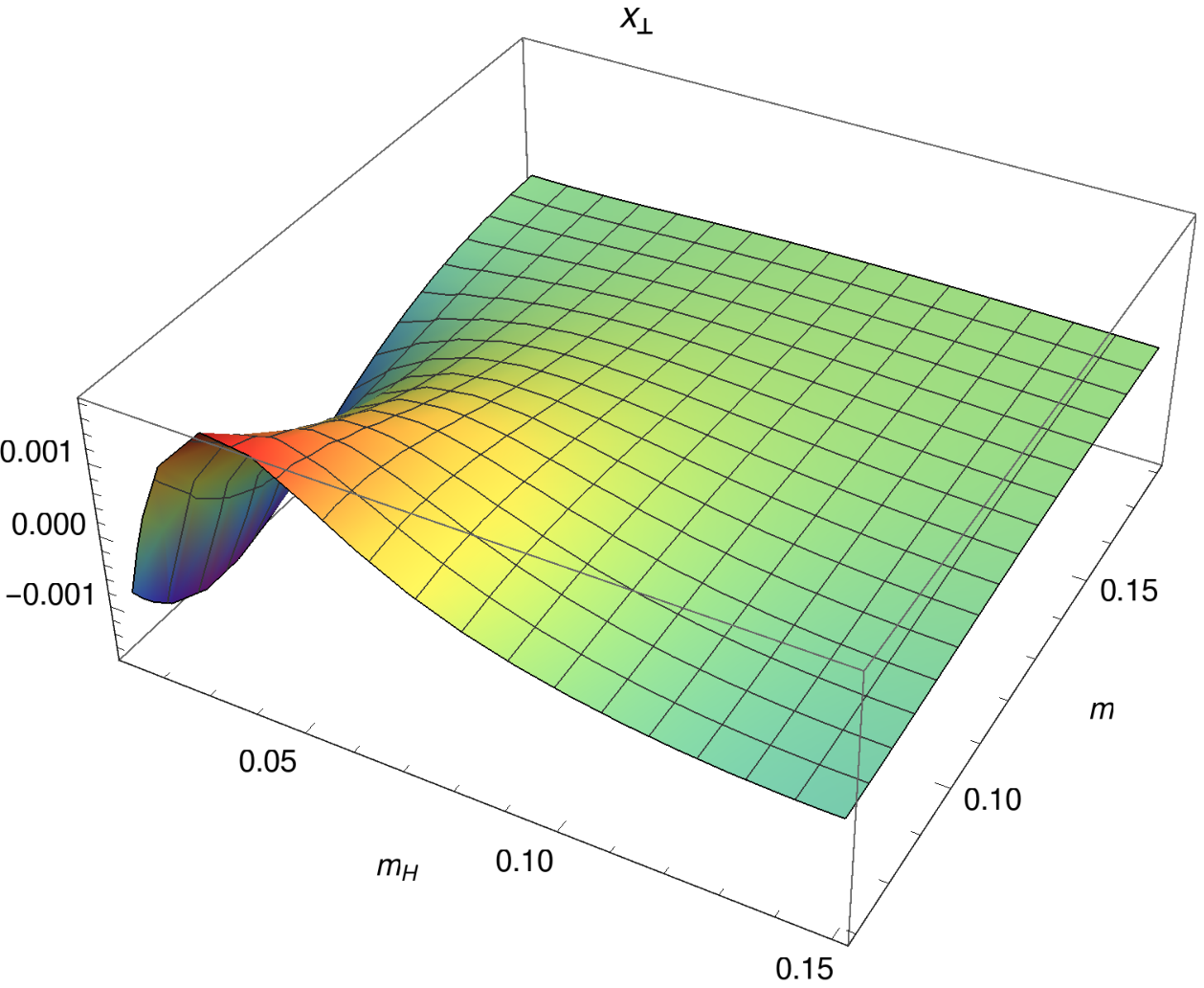}}
\end{center}
\caption{[Color online] Genuine spin-wave interaction manifesting itself in the parallel staggered susceptibility (LHS) and in the uniform
perpendicular susceptibility (RHS) of antiferromagnetic films in mutually orthogonal magnetic and staggered fields at the temperature
$T/2 \pi \rho_s = 0.1$.}
\label{figure7}
\end{figure}

To explore the parallel staggered susceptibility, we consider the ratio
\begin{equation}
\label{ratioSusParallel}
x_{\chi_{\parallel}}(T,H_s,H) = \frac{\chi^{\cal I}_s}{\chi^{[1]}_s T^{-1}} \, ,
\end{equation}
that measures strength and sign of the magnon-magnon interaction in the temperature-dependent piece of the parallel staggered
susceptibility relative to the dominant free Bose gas contribution. The result for the temperature $T/2 \pi \rho_s = 0.1$ is depicted on
the LHS of Fig.~\ref{figure7}. The spin-wave interaction is weak indeed and, depending on the region in parameter space defined by $m_H$
and $m$, the quantity $x_{\chi_{\parallel}}$ may take negative or positive values. But again, these two-loop interaction effects are rather
subtle.

We finally consider the uniform perpendicular susceptibility. Given the fact that the leading contribution $\chi^{[1]}$ takes both positive
and negative values in the parameter region of interest, it makes no sense to normalize the interaction part with respect to $\chi^{[1]} T$,
in analogy to $x_{\chi_{\parallel}}$, Eq.~(\ref{ratioSusParallel}). Rather, in order to avoid this artificial singularity, we superimpose the
zero-temperature contribution $\chi_{\perp}(0,H_s,H)$ and introduce the ratio
\begin{equation}
x_{\chi_{\perp}}(T,H_s,H) = \frac{\chi^{\cal I} T^2}{\chi_{\perp}(0,H_s,H) + \chi^{[1]} T} \, ,
\end{equation}
whose denominator always is positive. The quantity $x_{\chi_{\perp}}$ is plotted on the RHS of Fig.~\ref{figure7} for the temperature
$T/2 \pi \rho_s = 0.1$. Once again one observes that the magnon-magnon interaction is weak and that the quantity $x_{\chi_{\perp}}$ may take
positive or negative values.

\section{Conclusions}
\label{conclusions}

Using effective field theory, we have systematically addressed the question of how the magnon-magnon interaction manifests itself in
antiferromagnetic films in mutually perpendicular staggered and magnetic fields at finite temperature. To isolate the piece originating
from noninteracting magnons, we first had to evaluate the two-point functions and the respective dispersion relations for the two spin-wave
branches up to one-loop order. Within this dressed magnon picture, we then extracted the genuine spin-wave interaction piece in the
two-loop free energy density.

As it turns out, the magnon-magnon interaction in thermodynamic quantities is weak. In the pressure the genuine magnon-magnon interaction
is mainly repulsive but becomes attractive in weaker staggered and magnetic fields. As we illustrated in various figures, the genuine
spin-wave interaction, depending on the actual strength of the magnetic and staggered field, may lead to an increase of both the order
parameter and the magnetization at finite temperatures. This behavior is rather counterintuitive. On the other hand, the one-loop
contribution -- both in the order parameter and the magnetization -- is negative: if temperature increases (while keeping magnetic and
staggered fields fixed), order parameter and magnetization drop since the spins are perturbed thermally.

The parallel staggered and the uniform perpendicular susceptibilities exhibit similar characteristics: the role of the magnon-magnon
interaction in these observables is faint, and the effects induced at finite temperature are rather subtle. At zero temperature, the
parallel staggered susceptibility diverges in the double limit $\{H_s, H \} \to 0$, while the uniform perpendicular susceptibility tends to
a finite value that is fixed by the helicity modulus and the spin-wave velocity (hydrodynamic relation). Very precise loop-cluster results
for the helicity modulus and spin-wave velocity, available for the square and honeycomb lattice (and for $S=1/2$), allowed us to extract
accurate values for the uniform perpendicular susceptibility at $T$=0.

To realistically describe antiferromagnetic films as they are probed in experiments, additional types of interactions -- going beyond the
simple Heisenberg model -- would have to be considered as well. While this is in principle perfectly feasible within effective Lagrangian
field theory, here we restricted ourselves to the simple exchange interaction picture. Therefore we do not claim that the series derived
here do describe real antiferromagnetic samples in all details. Nevertheless, the basic characteristics of antiferromagnetic films,
subjected to mutually orthogonal magnetic and staggered fields, are captured adequately. Much like the many other references available on
the subject, the present investigation is a theoretical one, but for the first time it is fully systematic. Still, our effective field
theory predictions could be tested against numerical simulations also based on the "clean" Heisenberg Hamiltonian.

\end{document}